\documentclass[11pt]{article}
\pdfoutput=1

\usepackage{cite}
\usepackage{booktabs}
\usepackage[english]{babel}
\usepackage{amsmath,amssymb,amsbsy,amstext, amsthm, simplewick}
\usepackage{hyperref}
\usepackage{graphicx}
\usepackage{amsfonts}
\usepackage{amssymb}
\usepackage[small]{caption}
\usepackage{upgreek}
\usepackage[svgnames,dvipsnames,x11names,table]{xcolor}
\usepackage{multirow}
\usepackage{geometry}
\usepackage[hang,flushmargin]{footmisc}
\usepackage{bm}
\usepackage{braket}
\usepackage{subcaption}
\usepackage{mathtools}
\usepackage{setspace}
\usepackage{cleveref}
\usepackage{comment}
\usepackage{scalerel}
\usepackage[normalem]{ulem}
\usepackage{slashed}
\usepackage{enumitem}
\usepackage{dsfont}
\usepackage{tikz}
\usepackage{feynmp-auto}
\usetikzlibrary{decorations.markings}
\usetikzlibrary{shapes.misc}

\makeatletter
\g@addto@macro\bfseries{\boldmath}
\makeatother

\hypersetup{
    colorlinks=true,
    linkcolor={red!50!black},
    citecolor={blue!50!black},
    urlcolor={blue!80!black}
}

\usepackage{colortbl}

\makeatletter
\newlength{\apb@width}
\newcommand{\autoparbox}[2][c]{\settowidth{\apb@width}{#2}\parbox[#1]{\apb@width}{#2}}

\makeatother

\definecolor{lightgray}{gray}{0.9}

\usepackage[framemethod=default]{mdframed}
\newmdenv[skipabove=7pt,
skipbelow=7pt,
rightline=false,
leftline=false,
topline=false,
bottomline=false,
backgroundcolor=gray!10,
linecolor=gray,
innerleftmargin=5pt,
innerrightmargin=5pt,
innertopmargin=5pt,
innerbottommargin=5pt,
leftmargin=0cm,
rightmargin=0cm,
linewidth=4pt]{eBox}

\usepackage[most]{tcolorbox}
\tcbset{colback=white, colframe=black,
        highlight math style= {enhanced, %<-- needed for the ?remember? options
            colframe=red,colback=red!10!white,boxsep=0pt}
        }
\definecolor{light-gray}{gray}{0.95}

\crefname{table}{Table}{Tables}
\crefname{equation}{Eq.}{Eqs.}
\crefname{appendix}{App.}{Apps.}
\crefname{section}{Sec.}{Secs.}
\crefname{figure}{Fig.}{Figs.}

\numberwithin{equation}{section}

\def\beq{\begin{equation}}
\def\eeq{\end{equation}}

\def\bea{\begin{eqnarray}}
\def\eea{\end{eqnarray}}

\def\D{\mathcal{D}}
\def\d{{\rm d}}

\def\beq{\begin{equation}}
\def\eeq{\end{equation}}
\def\bea{\begin{eqnarray}}
\def\eea{\end{eqnarray}}

\def\D{{\cal D}}
\def\L{{\cal L}}

\def\d{{\rm d}}

\def\O{{\cal O}}

\def\d{{\rm d}}

\def\k{{\vec{\scaleto{k}{7pt}}}}

\def\x{{\vec x}}

\DeclareRobustCommand{\SkipTocEntry}[4]{}

\definecolor{colorTC}{rgb}{.2,.7,.2}

\definecolor{acolor}{rgb}{0.4, 0.2, 0.4}

\definecolor{blue3}{RGB}{31, 119, 180}
\definecolor{red3}{RGB}{	214, 39, 40}
\definecolor{orange3}{RGB}{255, 127, 14}
\definecolor{green3}{RGB}{44, 160, 44}

\begin{document}

\begin{titlepage}
\setcounter{page}{1} \baselineskip=15.5pt
\thispagestyle{empty}
$\quad$
\vskip 70 pt

\begin{center}
{\fontsize{20.74}{24} \bf Effective Field Theory and In-In Correlators}
\end{center}

\vskip 20pt
\begin{center}
\noindent
{\fontsize{12}{18}\selectfont Daniel Green and Guanhao Sun}
\end{center}

\begin{center}
\vskip 4pt
\textit{{\small Department of Physics, University of California at San Diego,  La Jolla, CA 92093, USA}}

\end{center}

%=========================================
\vspace{0.4cm}
 \begin{center}{\bf Abstract}
 \end{center}

\noindent

The predictions of inflation are usually defined in terms of equal time in-in correlation functions in an accelerating cosmological background. These same observables exist for quantum field theory in other spacetimes, including flat space. In this paper, we will explore how the Wilsonian renormalization group (RG) and effective field theory (EFT) apply to these observables in both flat and de Sitter space. Specifically, we show that matching the short- and long-distance calculations requires additional terms localized at the time of the measurement that are not captured by the effective action of the EFT. These additional terms only correct the local and semi-local terms in the EFT correlators. In flat space, we give an explicit demonstration by matching in-in correlators of light scalars interacting with a heavy field with the EFT result. We then show how these additional terms arise generically via exact RG. We also compare these explicit results in flat space with the corresponding theory in de Sitter and show that the local terms typically redshift away. Our results are closely related to momentum space entanglement that arises from tracing over short-wavelength modes.

\end{titlepage}

\setcounter{page}{2}

\restoregeometry

\begin{spacing}{1.2}
\newpage
\setcounter{tocdepth}{2}
\tableofcontents
\end{spacing}

\setstretch{1.1}
\newpage

\section{Introduction}

One of the central challenges in cosmology is that we only observe one universe. We do not perform experiments, or directly control the amount of data relevant to understanding our cosmic origins. The statistics of the primordial density fluctuations is one very important window into the very early Universe~\cite{Baumann:2018muz,Green:2022bre,Baumann:2022mni} that informs our understanding of the inflationary epoch or alternatives thereof~\cite{Achucarro:2022qrl,Flauger:2022hie}. Unfortunately, the space of models is much vaster than the volume of data itself. As such, we need theoretical input to help translate observations into insights~\cite{Green:2022hhj}.

In recent years, a host of techniques from quantum field theory in flat space have been enormously powerful in organizing our cosmological predictions. The development of the effective theory of inflation~\cite{Cheung:2007st,Cheung:2007sv,Senatore:2010wk,Baumann:2011su} has reduced the predictions of a wide range of models down to a few Wilson coefficients that can be constrained or measured~\cite{Senatore:2009gt,WMAP:2012fli,Planck:2019kim}. Symmetries of the adiabatic modes~\cite{Weinberg:2003sw} have also provided a sharp distinction between the predictions of single and multifield inflation~\cite{Maldacena:2002vr,Creminelli:2004yq}. More recently, the relationship between cosmological correlators and scattering amplitudes~\cite{Benincasa:2022gtd} has revealed that much of the structure of cosmological correlators has its origins in the dynamics of flat space and serves as an important input to the cosmological bootstrap~\cite{Arkani-Hamed:2018kmz,Baumann:2022jpr}.

Inspired by the bootstrap approach, there has been growing interest in the relationship between flat space and cosmological observables~\cite{Baumann:2015nta,Arkani-Hamed:2017fdk,Pajer:2020wnj,Goodhew:2020hob,Grall:2020ibl,Stefanyszyn:2020kay,Meltzer:2021zin,Melville:2021lst,Baumann:2021fxj,Goodhew:2021oqg,Cabass:2022jda,Albayrak:2023hie,Du:2024hol,Goodhew:2024eup,Thavanesan:2025kyc}. In-in correlation functions~\cite{Weinberg:2005vy} and/or the wavefunction of the universe~\cite{Maldacena:2002vr,Anninos:2014lwa} have received renewed attention in flat space, where their structure is remarkably similar to their cosmological avatars~\cite{Benincasa:2018ssx,Salcedo:2022aal}. As our understanding of quantum field theory in flat space is significantly more advanced, identifying known properties of physics in flat space via these observables is a compelling stepping stone to understanding the space of consistent cosmological models.

In this paper, we will explore the relationship between
EFTs defined by matching scattering amplitudes, in-in correlators, and wavefunction coefficients. For scattering amplitudes (in-out correlators), all information about short-distance theory is encoded in the coupling constants of the effective action. In-in correlators and the wavefunction have an additional complication due to the fact that they are defined at a preferred time, which breaks time translations and hence energy conservation. As a consequence, matching the in-in correlators in the EFT with the UV requires the addition of local terms to the EFT\footnote{A similar result was already noted for Wilsonian RG in the bulk of AdS, although the origin of the extra terms is slightly different in detail~\cite{Heemskerk:2010hk,Faulkner:2010jy}.}. These boundary terms in the action alter the statistics of the fields through local and semi-local terms in the correlations. However, they do not alter the residues of the total energy poles, as expected from the relationship between scattering amplitudes and in-in correlators~\cite{Maldacena:2011nz,Raju:2012zr}. We show this explicitly in the case of integrating out a heavy field and identify an infinite series of terms that are not captured by the effective action alone. These extra terms can be understood, through the uncertainty principle, as arising from the energy added to the system by the detectors in order to measure the field at a unique moment of time~\cite{Unruh:1983ms,Flauger:2013hra,Green:2020whw,Green:2022fwg}. 

The need for these corrections is not a peculiarity of the specific model but can be understood non-perturbatively to be the result of a typical RG flow. Using the framework of exact RG~\cite{Polchinski:1983gv}, we identify a single diagram that is responsible for the additional interaction terms that arise for in-in correlators. These corrections are local, but are not simply a local addition to the wavefunction.  Instead, integrating out short-wavelength modes turns a pure-state into a mixed-state so that the complete RG behavior is described by a density matrix~\cite{Goldman:2024cvx}. This analysis closely resembles the discussion of momentum space entanglement in EFT~\cite{Balasubramanian:2011wt,Hsu:2012gk,Flynn:2022tbj,Costa:2022bvs,Pelliconi:2023ojb,Hernandez-Cuenca:2024pey}.

In cosmological backgrounds, most of these subtleties vanish, as measurements take place when the universe is effectively classical (and not constrained by the uncertainty principle). As a result, the non-trivial boundary terms redshift away and we are left with local operators that commute with the correlator. In single field inflation, these local terms are fixed by a non-linearly realized symmetry that acts on the adiabatic mode~\cite{Pimentel:2013gza,Arkani-Hamed:2018kmz,Baumann:2019oyu,DuasoPueyo:2023kyh}. As a result, one is able to bootstrap the entire in-in correlator with the knowledge of the S-matrix~\cite{Pajer:2020wxk}. In contrast, these terms are allowed in multifield inflation (see e.g.~\cite{Lyth:2001nq,Zaldarriaga:2003my,Kumar:2019ebj,Wang:2022eop}) and is yet another indication of the qualitative difference between these two classes of models. In general, any additional (isocurvature) degree of freedom is subject to the ambiguity regarding local terms and thus is expected to generate some level of local non-Gaussianity unless it is tuned by hand. For the case of a massive field, we show explicitly that no addition terms arise in matching the UV and EFT descriptions.

This paper is organized as follows: In Section~\ref{sec:mass_flat}, we illustrate the relevant behavior in flat space by integrating out a single massive field interacting with a light field through tree level exchange. In Section~\ref{sec:ERG}, we derive the same behavior for a general scalar field theory using exact RG. In Section~\ref{sec:dS}, we apply our same model of light and heavy fields to cosmology and show that all the boundary terms vanish. We conclude in Section~\ref{sec:conclusions}. The main text is supplemented by three appendices: in Appendix~\ref{app:A}, we derive the boundary terms that are required by having a well-defined variational principle. In Appendix~\ref{app:B}, we derive the full expression for the exact RG, including the quadratic boundary terms. In Appendix~\ref{app:C}, we derive the in-in expression in de Sitter space using the Mellin representation.

We will use the following conventions throughout: $\vec k $, $\vec x$ denote vectors, while $k =|\vec k|$ or $x=|\x|$ are the lengths of a vector. We will often drop momentum conserving $\delta$-functions using the definition 
\beq
\langle ... \rangle = \langle ... \rangle' (2\pi)^3 \delta\left( \sum_i \k_i \right)\ .
\eeq
We will define the Fourier transform of the spatial coordinates of local field $\varphi(\x,t)$ by the replacement $\x \to \k$ or $\varphi(\k,t)$. We will use the variable $\k$ exclusively as wavenumbers to avoid confusion.

\section{In-In Correlators and EFT in Flat Space } \label{sec:mass_flat}
Let us start with an explicit example in flat spacetime. Consider a system whose Lagrangian is given by 
\begin{equation}
    S = \int d^4 x \left[ -\frac{1}{2}\partial_\mu\phi\partial^\mu\phi - \frac{1}{2}\partial_\mu\sigma\partial^\mu\sigma - \frac{1}{2}M^2 \sigma^2 - \lambda \phi^2 \sigma \right] \ ,
\end{equation}
where $\phi$, $\sigma$ are real scalar fields. We will compute correlators of this theory both by direct in-in calculation and through the wavefunction formalism. We will then integrate out the heavy field $\sigma$ and match the correlators in the EFT for $\phi$ with the UV calculations, paying particular attention to terms not captured by the effective action derived from in-out calculations. 

\subsection{Direct Calculations}

The in-in formalism is typically defined directly as a perturbative framework to calculate correlators in terms of canonically quantized fields and the interaction Hamiltonian. The interaction picture operators for our model are given by
\begin{align}
    \phi(\x,t) &=\int \frac{d^3 k}{(2\pi)^3} f^*_k(t)\hat a^{(\phi)}_{\vec k} + f_k(t) \hat a^\dagger{}^{(\phi)}_{-\vec k} \ , \\
    \sigma(\x,t) &=\int \frac{d^3 k}{(2\pi)^3} g^*_k(t)\hat a^{(\sigma)}_{\vec k} + g_k(t) \hat a^\dagger{}^{(\sigma)}_{-\vec k} \ ,
\end{align}
where
\begin{align}
    f_k(t) = \frac{e^{i k (t-t_0)}}{\sqrt{2 k}} , \qquad \qquad 
    g_k(t) = \frac{e^{i E_k (t-t_0)}}{\sqrt{2 E_k}} \ ,
\end{align}
are the solutions to the free equations of motion, $E_k = \sqrt{k^2 + M^2}$, and the normalization is chosen so that the ladder operators satisfy
\begin{equation}
    [\hat a_{\vec k} , \hat a^{\dagger}_{\vec k '}] = (2\pi)^3 \delta^3 (\vec k - \vec k') \ .
\end{equation}
Given an interaction Hamiltonian, $H_I$, we use the standard nested commutator formula\footnote{The commutator form only applies when $i\epsilon = 0$. At tree level, the usual $i\epsilon$ prescription is the same as evaluating the indefinite integrals and dropping the contribution from $-\infty$.} for the in-in correlators \cite{Weinberg:2005vy}, 
\begin{equation}
    \langle \O(t_0) \rangle = \sum_{N=0}^\infty i^N \int_{-\infty}^{t_0} dt_N \int_{-\infty}^{t_N} \cdots \int_{-\infty}^{t_2} dt_1 \langle [H_I(t_1),[H_I(t_2),\cdots [H_I(t_N),\O(t_0)] \cdots ]] \rangle \ .
\end{equation}
In the rest of this section we set $t_0 = 0$ unless otherwise stated to simplify the expressions.

Given the Lagrangian, we have $H_I = - \int d^3x \; \L_I =  \lambda \int d^3x \; \phi^2 \sigma$. We will calculate the bispectrum $\langle \phi^2 \sigma \rangle$ and the trispectrum $\langle \phi^4 \rangle$. The first contribution to the bispectrum arises at linear order in $\lambda$, 
\begin{align}\label{eq:in-in_phi^2sigma}
    \langle \phi(\vec k_1, 0) \phi(\vec k_2, 0) \sigma(\vec k_3, 0) \rangle' =& i\lambda \int_{-\infty}^{0} dt \; 2 \langle [\phi(\vec k_1, t) \phi(\vec k_2, t) \sigma(\vec k_3, t), \phi(\vec k_1, 0) \phi(\vec k_2, 0) \sigma(\vec k_3, 0)]\rangle \nonumber \\
    =& -\frac{\lambda}{2k_1 k_2 E_{\vec k_1 + \vec k_2}\left(k_1+k_2 + E_{\vec k_1 + \vec k_2} \right)} + \mathcal{O}(\lambda^2) \ ,
\end{align}
where $E_{\vec k} = \sqrt{\vec k \cdot \vec k + M^2}$, and we included a symmetry factor of 2 from the exchange of the two $\phi$'s in $H_I$. Notice that this bispectrum is non-zero and is only polynomially suppressed in terms of the mass of the heavy field. This is a hint that we probably should not simply set the value of the heavy field to zero at the boundary $t=0$ when calculating low-energy correlators. 

Next, we calculate the contribution of exchange diagrams to the trispectrum $\langle \phi^4 \rangle$. To avoid clustering, we will be using the notations $\phi_{i}(t) = \phi(\vec k_i, t)$, and $\bar \phi_i = \phi(\vec k_i, 0)$. Then, at leading order in $\lambda$, 
\begin{align}\label{eq:in-in_phi^4}
    \langle \bar\phi_1 \bar\phi_2 \bar\phi_3 \bar\phi_4 \rangle ' \supset& \;\;\;\;\, 4 i^2 \lambda^2 \int_{-\infty}^0 dt_2 \int_{-\infty}^{t_2} dt_1 \; \langle [ \phi_3\phi_4 \sigma_5(t_1),[\phi_1\phi_2\sigma_5(t_2), \bar\phi_1 \bar\phi_2 \bar\phi_3 \bar\phi_4]] \rangle \nonumber \\
    &+ 4 i^2 \lambda^2 \int_{-\infty}^0 dt_2 \int_{-\infty}^{t_2} dt_1 \; \langle [ \phi_1\phi_2 \sigma_5(t_1),[\phi_3\phi_4\sigma_5(t_2), \bar\phi_1 \bar\phi_2 \bar\phi_3 \bar\phi_4]] \rangle \nonumber \\
    &+ ``t"+``u" \nonumber \\
    =& \;\;\;\;\, \frac{\lambda^2}{4 k_1 k_2 k_3 k_4 E_{k_5}} \left( \frac{1}{(k_3 + k_4 + E_{k_5})k_T} + \frac{1}{(k_3 + k_4 + E_{k_5})(k_1+k_2-k_3-k_4)} \right) \nonumber \\
    &+ \frac{\lambda^2}{4 k_1 k_2 k_3 k_4 E_{k_5}} \left( \frac{1}{(k_1 + k_2 + E_{k_5})k_T} + \frac{1}{(k_1 + k_2 + E_{k_5})(k_3+k_4-k_1-k_2)} \right) \nonumber \\
    &+ ``t"+``u" \nonumber \\
    =& \frac{\lambda^2(E_{k_5} + k_T)}{ 2k_1 k_2 k_3 k_4 E_{k_5} (k_1 + k_2 + E_{k_5} )(k_3 + k_4 + E_{k_5} )k_T} + ``t"+``u".
\end{align}
Here we use $k_T = k_1 + k_2 + k_3 + k_4$ and $\vec k_5 = -\vec k_1 - \vec k_2 = -\vec k_3 - \vec k_4$ in the $``s"$-channel.  The $``t"$ and $``u"$ terms stand for the contributions where $k_5= - \k_1-\k_3$ and $k_5 =-\k_1-\k_4$ respectively. A curious property of this result is that, when expanded in large mass $M$, there are terms that are of odd powers in $M$. More explicitly,  
\begin{align}\label{eq:phi4_real_expansion}
    \frac{E_{k_5}+k_T}{E_{k_5}(E_{k_5} + k_{12})(E_{k_5} + k_{34})k_T} =& \;\;\;\;\, \frac{1}{M^2 k_T} - \frac{k_{12}k_{34}+ k_5^2 }{M^4 k_T} + \frac{k_{12}k_{34}}{M^5} \nonumber \\
    &+\frac{1}{M^6 k_T}\left( (k_5^2 + k_{12}k_{34})^2 - k_{12}k_{34} k_T^2 \right) + \mathcal{O}(\frac{1}{M^7}) \ ,
\end{align}
where we introduced the notation $k_{ij} = k_i + k_j$. This differs from the traditional EFT wisdom where we can only have $M$ in even powers. To understand these terms, it is helpful to write 
\begin{align} \label{eq:phi4_EFT_plus_boundary}
    &\frac{(E_{k_5} + k_T)}{ E_{k_5} (k_1 + k_2 + E_{k_5} )(k_3 + k_4 + E_{_5} )k_T} \nonumber \\
    =& \frac{1}{2k_T (E_{k_5}^2 - k_{12}^2)} + \frac{1}{2k_T (E_{k_5}^2 - k_{34}^2)} + \frac{-E_{k_5} k_T + 2 k_{12}k_{34}}{2E_{k_5} (E_{k_5}^2 - k_{12}^2) (E_{k_5}^2 - k_{34}^2)} \ ,
\end{align}
where the first two terms of the last line come from EFT-like terms, and the third term contains all the terms that are odd in $M$. Notice that the third term does not have a pole at $k_T=0$, so that terms of odd power in $M$ do not induce any unexpected contributions in the corresponding scattering amplitude. In fact, the residue of the total energy ($k_T=0$) pole can be exactly matched to the usual EFT contribution (the first term of Equation~(\ref{eq:in-in_S_EFT})). The third term contains all the information not encoded in the scattering amplitude (and hence in the conventional EFT). One can further split this term into even and odd in $1/M$ terms, namely 
\begin{align}
    &\frac{-E_{k_5} k_T + 2 k_{12}k_{34}}{2E_{k_5} (E_{k_5}^2 - k_{12}^2) (E_{k_5}^2 - k_{34}^2)} = \frac{- k_T }{ 2(E_{k_5}^2 - k_{12}^2) (E_{k_5}^2 - k_{34}^2)} + \frac{  k_{12}k_{34}}{E_{k_5} (E_{k_5}^2 - k_{12}^2) (E_{k_5}^2 - k_{34}^2)} \nonumber \\
    =& \frac{- k_T }{ 2(M^2 + k_5^2 - k_{12}^2) (M^2 + k_5^2 - k_{34}^2)} + \frac{  k_{12}k_{34}}{M\sqrt{1 + \frac{k_5^2}{M^2}} (M^2 + k_5^2 - k_{12}^2) (M^2 + k_5^2 - k_{34}^2)} \ . \label{eq:missing_terms}
\end{align}
The first part is manifestly even in $1/M$, while the second part is odd. This means that we have both even and odd $1/M$ corrections compared with naive EFT result. Notice that the odd terms lead to corrections with analytic structures of the form $k_{12} k_{34} / k_1 k_2 k_3 k_4$, which is analytic in two out of the four momenta. These terms are semi-local, corresponding to measurements with two points overlapping with each other in physical space, and we do not expect an EFT prescription to capture them correctly. We defer detailed discussions to section \ref{sec:EFT_and_Boundary_Terms}. 

In the rest of this section we will discuss this result with the help of the wavefunction method and see what kind of modifications to the traditional EFT, both as boundary terms and as semi-local terms, are needed to reproduce the necessary corrections. 

\subsection{Relation to the Wavefunction Method}
As promised, let us now turn to the calculation using the wavefunction of this system. The starting point is to realize that a correlator can be written as 
\begin{align}
    \langle \O(t=0) \rangle =& \int \D \bar \phi  \D \bar \sigma \; \langle 0 | \bar \phi, \bar \sigma \rangle \langle \bar \phi, \bar \sigma | \O |0\rangle \nonumber \\
    =& \int \D \bar \phi \D \bar \sigma \left( \int^{ \bar \phi, \bar\sigma} \D \phi' \D \sigma' \; e^{iS[\phi', \sigma']}\right)^* \O[\bar\phi, \bar\sigma] \left( \int^{ \bar \phi, \bar\sigma} \D \phi \D \sigma \; e^{iS[\phi, \sigma]}\right) \nonumber \\
    =& \int \D \bar \phi \D \bar \sigma \; |\Psi[\bar\phi, \bar\sigma]|^2 \; \O[\bar\phi, \bar\sigma] \ .
\end{align}
The wavefunction of the system is then defined by
\begin{equation}
    \Psi[\bar\phi, \bar\sigma] =  \int^{ \bar \phi, \bar\sigma} \D \phi \D \sigma \; e^{iS[\phi, \sigma]} \propto e^{iS_{\text{cl}}[\phi, \sigma]} \ ,
\end{equation}
where in the last step we have used the saddle point approximation. That is, the wavefunction of the system, to the leading order in saddle point approximation, is the exponential of the classical action, properly normalized. One then plug in the saddle point solution of the field equations into the action. To do this, it is convenient to write the solution to the classical equations of motion as 
 \begin{equation}
     \phi_{\text{cl}}(\vec x, t) = \int d^3x' \; K_{\phi}(t,\vec x, \vec x') \bar\phi(\vec x') + \int d^4 x' \; G_{\phi}(x,x') \frac{\delta S_{\text{int}}}{\delta \phi_{\text{cl}}(x')} \ ,
 \end{equation}
 with the propagators 
 \begin{align}
     K_{\phi, \sigma}(k,t) =& e^{i\omega_{\phi,\sigma}(k) t}  \\
     G_{\phi,\sigma}(k,t,t') =& \frac{i}{2\omega_{\phi,\sigma}(k)}\left( e^{i\omega_{\phi,\sigma}(k) (t - t')} \theta (t' - t) + e^{i\omega_{\phi,\sigma}(k) (t' - t)} \theta (t - t') - e^{i\omega_\phi(k)(t+t')}  \right) \ .
 \end{align}
 Here $\omega_\phi = k$ and $\omega_\sigma = E_k=\sqrt{M^2 + k^2}$ are the energies of the $\phi$ and $\sigma$ fields respectively. The bulk-to-bulk propagator $G$ is defined so that $G(x,x') = 0$ for either $t=0$ or $t' = 0$. This arrangement makes transparent the fact that $\phi(t=0) = \bar \phi$. 
 
Using this ansatz to evaluate the action and suppressing $\delta$-functions and $k$ integrals, we have to order $\lambda^2$,
\begin{align}
    \log \Psi \supset& -\frac{1}{2}k \bar \phi_{\vec k} \bar \phi_{-\vec k} -\frac{1}{2}E_{\vec k} \bar \sigma_{\vec k} \bar \sigma_{-\vec k} - \frac{\lambda}{k_1 + k_2 + E_{\vec k_1 + \vec k_2}}\bar \phi_{\vec k_1} \bar \phi_{\vec k_2} \bar \sigma_{\vec k_3} \nonumber \\
    &+\frac{2\lambda^2}{(k_1 + E_2 + k_5 )(k_3 + E_4 + k_5 )(k_1 + E_2 + k_3 + E_4 )} \bar \phi_{\vec k_1} \bar \sigma_{\vec k_2} \bar \phi_{\vec k_3}\bar \sigma_{\vec k_4}  \\
    &+ \frac{\lambda^2}{2} \frac{1}{(k_1 + k_2 + E_{k_5})(k_3 + k_4 + E_{k_5})(k_1 + k_2 + k_3 + k_4)} \bar \phi_{\vec k_1} \bar \phi_{\vec k_2} \bar \phi_{\vec k_3}\bar \phi_{\vec k_4} + ``t" + ``u"  \ , \nonumber 
\end{align}
where the $s$, $t$, and $u$ channels are defined as before. The functions of $\k_i$ multiplying the powers of $\bar \phi$ and $\bar \sigma$ are called the coefficients of the wavefunction. For any weakly coupled theory, we can write $\Psi$ as an expansion around the Gaussian wavefunction, 
\begin{align}
    \Psi[\bar\phi, \bar\sigma] =& \exp\left(-\frac{1}{2}\int \frac{d^3 k }{(2\pi)^3} (\psi_{\phi^2}\bar \phi_{\vec k} \bar \phi_{-\vec k} + \psi_{\sigma^2} \bar \sigma_{\vec k} \bar \sigma_{-\vec k} ) \right) \nonumber \\
    &\times \bigg[ 1 + \int \frac{d^3k_1 d^3k_2 d^3k_3}{(2\pi)^6}\; \psi_{\phi_1\phi_2 \sigma_3} \bar\phi_1\bar\phi_2\bar\sigma_3 \; \delta^3(\sum_{i=1}^3 \k_i)  \\
    &\;\;\;\;+ \frac{1}{2} \int \frac{d^3k_1 d^3k_2 d^3k_3 d^3k_4 d^3k_5 d^3k_6}{(2\pi)^{12}} \; \psi_{\phi_1\phi_2 \sigma_3} \bar\phi_1 \bar\phi_2 \bar\sigma_3 \; \psi_{\phi_4\phi_5 \sigma_6} \bar\phi_4\bar\phi_5\bar\sigma_6 \; \delta^3(\sum_{i=1}^3 \k_i) \delta^3(\sum_{j=4}^6 \k_j) \nonumber \\
    &\;\;\;\;+ \int \frac{d^3k_1 d^3k_2 d^3k_3 d^3k_4 }{(2\pi)^{9}}\;  \psi_{\phi_1\phi_2\phi_3\phi_4} \bar\phi_1 \bar\phi_2 \bar\phi_3 \bar\phi_4 \; \delta^3(\sum_{i=1}^4 \k_i) + \cdots \bigg] \nonumber \ .
\end{align}
From now on we use $\psi$ to represent the coefficients of the wavefunction\footnote{Note that the sign convention is different between the quadratic (Gaussian) and higher order terms.}, with the $\k$ dependence of these coefficients given in the subscripts. From the explicit calculation, all of the coefficients in our example are real\footnote{This is specific to our simple example in flat spacetime. It is not true in general.}. We also require that $\bar\phi$ and $\bar \sigma$ are real. This simplifies our calculations and gives us 
\begin{equation} \label{eq:wavefunction_phi^2sigma}
    \langle \bar\phi_1 \bar\phi_2 \bar\sigma_3 \rangle' = 2\times \frac{2\psi_{\phi_1 \phi_2 \sigma_3}}{8\psi_{\phi^2_1} \psi_{\phi^2_2} \psi_{
    \sigma^2_3}} = - \frac{\lambda}{2k_1 k_2 E_3(k_{12}+E_3)} \ ,
\end{equation}
matching exactly the in-in calculation (\ref{eq:in-in_phi^2sigma}). Here the first factor of 2 is a symmetry factor, and $2\psi_{\phi_1 \phi_2 \sigma_3}$ comes from $\psi_{\phi_1 \phi_2 \sigma_3}+\psi^*_{\phi_1 \phi_2 \sigma_3}$. 

The fact that this correlator is non-zero, despite the large mass for $\sigma$, arises from the energy of the detector that is needed to localize this correlator at a specific time~\cite{Unruh:1983ms,Flauger:2013hra,Green:2020whw,Green:2022fwg}. Still we note that the expansion in $k_3 \ll M$ yields an analytic function of $\k_3$ and therefore is localized at coincident points. This is consistent with the expectation that $\sigma$ is in the vacuum (unexcited) until the time of the measurement and therefore cannot generate long-range correlations.

\begin{figure}
    \centering
    \begin{fmffile}{r-l}
        \begin{fmfgraph*}(75,45)
            \fmfstraight
            \fmfleft{i0,i3,i6}
            \fmfright{o0,o3,o6}
            \fmf{vanilla}{i3,o3}
            \fmf{vanilla}{i3,v1}
            \fmf{vanilla}{v1,v2}
            \fmf{vanilla}{v2,v3}
            \fmf{vanilla}{v3,v4}
            \fmf{vanilla}{v4,v5}
            \fmf{vanilla}{v5,o3}
            \fmf{phantom,tension=2}{i6,v6}
            \fmf{phantom}{v6,o6}
            \fmf{phantom,tension=2}{o0,v7}
            \fmf{phantom}{i0,v7}
            \fmffreeze
            \fmf{vanilla}{v1,v6,v2}
            \fmf{dashes}{v3,v6}
            \fmf{vanilla}{v5,v7,v4}
            \fmf{dashes}{v3,v7}
        \end{fmfgraph*}
    \end{fmffile}
    \raisebox{7mm}{+}
    \begin{fmffile}{l-r}
        \begin{fmfgraph*}(75,45)
            \fmfstraight
            \fmfleft{i0,i3,i6}
            \fmfright{o0,o3,o6}
            \fmf{vanilla}{i3,o3}
            \fmf{vanilla}{i3,v1}
            \fmf{vanilla}{v1,v2}
            \fmf{vanilla}{v2,v3}
            \fmf{vanilla}{v3,v4}
            \fmf{vanilla}{v4,v5}
            \fmf{vanilla}{v5,o3}
            \fmf{phantom,tension=2}{i0,v6}
            \fmf{phantom}{v6,o0}
            \fmf{phantom,tension=2}{o6,v7}
            \fmf{phantom}{i6,v7}
            \fmffreeze
            \fmf{vanilla}{v1,v6,v2}
            \fmf{dashes}{v3,v6}
            \fmf{vanilla}{v5,v7,v4}
            \fmf{dashes}{v3,v7}
        \end{fmfgraph*}
    \end{fmffile}
    \raisebox{7mm}{+}
    \begin{fmffile}{ll}
        \begin{fmfgraph*}(75,45)
            \fmfstraight
            \fmfleft{i0,i3,i6}
            \fmfright{o0,o3,o6}
            \fmf{vanilla}{i3,o3}
            \fmf{vanilla}{i3,v1}
            \fmf{vanilla}{v1,v2}
            \fmf{vanilla}{v2,v3}
            \fmf{vanilla}{v3,v4}
            \fmf{vanilla}{v4,v5}
            \fmf{vanilla}{v5,o3}
            \fmf{phantom,tension=2}{i0,v6}
            \fmf{phantom}{v6,o0}
            \fmf{phantom,tension=2}{o0,v7}
            \fmf{phantom}{i0,v7}
            \fmffreeze
            \fmf{vanilla}{v1,v6,v2}
            \fmf{dashes}{v3,v6}
            \fmf{vanilla}{v5,v7,v4}
            \fmf{dashes}{v3,v7}
        \end{fmfgraph*}
    \end{fmffile}
    \raisebox{7mm}{+}
    \begin{fmffile}{rr}
        \begin{fmfgraph*}(75,45)
            \fmfstraight
            \fmfleft{i0,i3,i6}
            \fmfright{o0,o3,o6}
            \fmf{vanilla}{i3,o3}
            \fmf{vanilla}{i3,v1}
            \fmf{vanilla}{v1,v2}
            \fmf{vanilla}{v2,v3}
            \fmf{vanilla}{v3,v4}
            \fmf{vanilla}{v4,v5}
            \fmf{vanilla}{v5,o3}
            \fmf{phantom,tension=2}{i6,v6}
            \fmf{phantom}{v6,o6}
            \fmf{phantom,tension=2}{o6,v7}
            \fmf{phantom}{i6,v7}
            \fmffreeze
            \fmf{vanilla}{v1,v6,v2}
            \fmf{dashes}{v3,v6}
            \fmf{vanilla}{v5,v7,v4}
            \fmf{dashes}{v3,v7}
        \end{fmfgraph*}
    \end{fmffile}
    \caption{Diagrams that correspond to the terms that combine into the second term of equation (\ref{eq:phi^4_wavefunction_full}). Later we will see how these terms show up in the exact RG calculation. Vertices above and below the horizontal $t=0$ line represents left and right vertices respectively. Solid lines are $\phi$ and dashed lines are $\sigma$ fields.}
    \label{fig:tree_exchange_semilocal}
\end{figure}
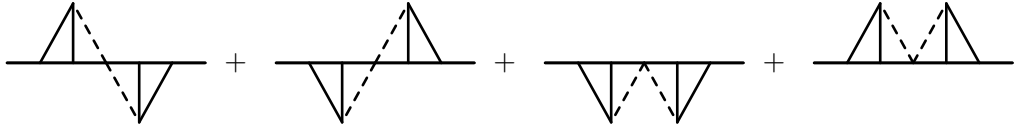

The trispectrum is more involved but still quite tractable. In the $s$ channel, it is given by
\begin{align} \label{eq:phi^4_wavefunction_full}
    \langle \bar\phi_1 \bar\phi_2 \bar\phi_3 \bar\phi_4 \rangle \supset& 8 \times \left( \frac{2\psi_{\phi_1 \phi_2 \phi_3 \phi_4}}{16 \psi_{\phi_1^2} \psi_{\phi_2^2} \psi_{\phi_3^2} \psi_{\phi_4^2}} + \frac{4 \psi_{\phi_1 \phi_2 \sigma_5} \psi_{\phi_3 \phi_4 \sigma_5} }{2\times32 \psi_{\phi_1^2} \psi_{\phi_2^2} \psi_{\phi_3^2} \psi_{\phi_4^2} \psi_{\sigma_5^2}}\right) \nonumber \\
    =& \frac{\lambda^2 (E_5+k_T)}{2k_1 k_2 k_3 k_4 E_5 k_T (k_{12}+E_5)(k_{34}+E_5)} \ .
\end{align}
Again, the factor of $8$ is a symmetry factor. This matches with the direct in-in calculation (\ref{eq:in-in_phi^4}) exactly. We can interpret the first term as the contact diagram and the second term as a summation of several ``exchange" diagrams. In fact, the second term is a sum of four terms that are, schematically, $\psi_{125}\psi_{345}+\psi^*_{125}\psi^*_{345}+\psi^*_{125}\psi_{345}+\psi_{125}\psi^*_{345}$. They combine into a single term because all these $\psi$'s are real in our example. Diagrammatically, these can be represented by four tree level diagrams with the heavy field propagator reaching the boundary, shown in Fig. \ref{fig:tree_exchange_semilocal}. We are using the conventions in, e.g., \cite{Giddings:2010ui}, to represent the left and right vertices by putting them above and below the $t=0$ line, respectively. The first two of them are basically exchange diagrams that connects the two sides of the path integral, while the other two are exchange diagrams on the same side, but with the heavy field propagator touching the boundary. They will show up again in our EFT calculations to come. 

Lastly, notice that both the contact and the exchange parts contain terms odd in $1/M$. For the contact part, 
\begin{align}
    2\psi_{\phi_1 \phi_2 \phi_3 \phi_4} =& \frac{\lambda^2}{(k_1 + k_2 + E_{k_5})(k_3+k_4+E_{k_5})k_T} = \frac{\lambda^2(E_{k_5}^2 - k_T E_{k_5} + k_{12}k_{34})}{(E_{k_5}^2 - k_{12}^2)(E_{k_5}^2 - k_{34}^2)k_T} \nonumber \\
    =&\frac{\lambda^2(M^2 + k_5^2 + k_{12}k_{34})}{(M^2 + k_5^2 - k_{12}^2)(M^2 + k_5^2 - k_{34}^2)k_T} - \frac{\lambda^2 M \sqrt{1+\frac{k_5^2}{M^2}}}{(M^2 + k_5^2 - k_{12}^2)(M^2 + k_5^2 - k_{34}^2)} \ ,
\end{align}
and for the exchange part, 
\begin{align}
    \frac{ \psi_{\phi_1 \phi_2 \sigma_5} \psi_{\phi_3 \phi_4 \sigma_5} }{ \psi_{\sigma_5^2}}=& \frac{\lambda^2}{(k_{12}+E_{k_5})(k_{34}+E_{k_5})E_{k_5}} = \frac{\lambda^2(E_{k_5}^2 - k_T E_{k_5} + k_{12}k_{34})}{(E_{k_5}^2 - k_{12}^2)(E_{k_5}^2 - k_{34}^2)E_{k_5}} \\
    =&\frac{\lambda^2(M^2 + k_5^2 + k_{12}k_{34})(1+\frac{k_5^2}{M^2})^{-1/2}}{M (M^2 + k_5^2 - k_{12}^2)(M^2 + k_5^2 - k_{34}^2)} - \frac{\lambda^2 k_T }{(M^2 + k_5^2 - k_{12}^2)(M^2 + k_5^2 - k_{34}^2)} \nonumber  \ .
\end{align}
Comparing to (\ref{eq:phi4_EFT_plus_boundary}), we see that both the contact and the exchange diagrams contribute to both even and odd in $1/M$ corrections to EFT. 

\subsection{EFT, Boundary Terms, and Entanglement} \label{sec:EFT_and_Boundary_Terms}
One of the advantages of the wavefunction method is to make the process of integrating out the heavy field transparent, as we can simply carry out the integral in $\bar\sigma$ in the path integral. This is understood as a trace over the heavy field. Specifically, the UV wavefunction is given by $\Psi[\phi,\sigma]$ and therefore integrating out $\sigma$ corresponds to the reduced density matrix
\beq
\rho[\bar \phi,\bar \phi'] = {\rm Tr}_{\bar\sigma}  |\bar\phi,\bar\sigma \rangle  \langle \bar\phi',\bar\sigma| = \int {\cal D} \sigma \Psi[\bar\phi,\bar\sigma] \Psi^*[\bar\phi',\bar \sigma] \ .
\eeq
A priori, it is not obvious that integrating out $\sigma$ is necessarily described in terms of a (reduced) density matrix. For example, we can rewrite the wavefunction, by completing the square, as 
\begin{align}
    \log \Psi \supset& -\frac{1}{2}k \bar \phi_{\vec k} \bar \phi_{-\vec k} -\frac{1}{2}E_{\vec k} \left(  \bar \sigma_{\vec k}  + \frac{\lambda}{k_{12}+ E_{\vec k}}\bar \phi_{\vec k_1} \bar \phi_{\vec k_2} \right)\left(  \bar \sigma_{-\vec k}  + \frac{\lambda}{k_{34}+ E_{-\vec k}}\bar \phi_{\vec k_3} \bar \phi_{\vec k_4} \right)\nonumber \\
    &+ \frac{\lambda^2}{2(k_1 + k_2 + E_k)(k_3 + k_4 + E_k)E_k} \bar \phi_{\vec k_1} \bar \phi_{\vec k_2} \bar \phi_{\vec k_3}\bar \phi_{\vec k_4} + ``t" + ``u"\nonumber \\
    &+\frac{2\lambda^2}{(k_1 + E_2 + k )(k_3 + E_4 + k )(k_1 + E_2 + k_3 + E_4 )} \bar \phi_{\vec k_1} \bar \sigma_{\vec k_2} \bar \phi_{\vec k_3}\bar \sigma_{\vec k_4} \nonumber \\
    &+ \frac{\lambda^2}{2} \frac{1}{(k_1 + k_2 + E_k)(k_3 + k_4 + E_k)(k_1 + k_2 + k_3 + k_4)} \bar \phi_{\vec k_1} \bar \phi_{\vec k_2} \bar \phi_{\vec k_3}\bar \phi_{\vec k_4} + ``t" + ``u" \nonumber \\
    &+ \cdots
\end{align}
One might imagine integrating out $\sigma$ directly at the wavefunction level to arrive at an effective $\Psi_{\rm EFT}[\phi]$ that is still a pure state. However, we can still compute the density matrix, in which case we have 
\begin{align}
&\rho[\bar\phi,\bar\phi'] =  \int {\cal D} \sigma  \exp\Bigg[-\frac{k}{2}\left(\bar \phi_{\vec k} \bar \phi_{-\vec k}+\bar \phi'_{\vec k} \bar \phi'_{-\vec k} \right) \nonumber \\
&-E_{\vec k} \left(  \bar \sigma_{\vec k}  +\frac{1}{2} \frac{\lambda}{k_{12}+ E_{\vec k}}(\bar \phi_{\vec k_1} \bar \phi_{\vec k_2} +\bar \phi'_{\vec k_1} \bar \phi'_{\vec k_2} )\right)\left(  \bar \sigma_{-\vec k}  + \frac{\lambda}{k_{34}+ E_{-\vec k}}(\bar \phi_{\vec k_3}\bar \phi_{\vec k_4} +\bar \phi'_{\vec k_3}\bar \phi'_{\vec k_4})\right) \\
&+ \frac{\lambda^2}{2(k_1 + k_2 + E_k)(k_3 + k_4 + E_k)E_k}  (\bar \phi_{\vec k_1} \bar \phi_{\vec k_2}+\phi'_{\vec k_1} \bar \phi'_{\vec k_2})(\bar \phi_{\vec k_3}\bar \phi_{\vec k_4}+\bar \phi'_{\vec k_3}\bar \phi'_{\vec k_4} )+ ``t" + ``u" + \ldots\Bigg] \nonumber \ .
\end{align}
We can perform the integral over $\bar\sigma$ by a change of variables,
\beq
\bar \sigma'_{\vec k} = \bar \sigma_{\vec k}  +\frac{1}{2} \frac{\lambda}{k_{12}+ E_{\vec k}}(\bar \phi_{\vec k_1} \bar \phi_{\vec k_2} +\bar \phi'_{\vec k_1} \bar \phi'_{\vec k_2} ) \ ,
\eeq
to arrive at a Gaussian integral for the second line. From the cross-terms of the form $\bar \phi^2 \bar \phi'{}^2$ that remain, we see this is not equivalent to a pure state defined by some $\Psi_{\rm EFT}$. Interestingly, this is not manifest from the expression for the in-in correlators directly, where
\beq
\langle {\rm in} | {\cal O}[\phi] |{\rm in} \rangle = \int D \bar \phi \, {\cal O}[\bar \phi] \, \rho[\bar \phi,\bar \phi] \ ,
\eeq
as the expression for the diagonal entries of the density matrix, with $\bar \phi = \bar \phi'$, could be represented by an effective wavefunction. To avoid confusion, we use $r_{{\cal O}_1 ..{\cal O}_n}$ to represent terms in $\log \rho$ associated with the operator in the subscript (in analogy with the wavefunction coefficients $\psi_{\phi_1.. \phi_n}$ for example). This notation is introduced to emphasize that these coefficients are not necessary wavefunction coefficients, but may only appear in the reduced density matrix. Concretely, the terms relevant to the in-in correlators in the EFT include 
\begin{equation}
    \log \rho[\bar\phi, \bar\phi] = -\frac{1}{2} r_{\phi^2} \bar\phi_{\k} \bar\phi_{-\k} + r_{\phi^4} \bar\phi_1\bar\phi_2\bar\phi_3\bar\phi_4 + \cdots
\end{equation}
The diagonal terms are all we will need for matching correlators, but conceptually the off-diagonal terms in the density matrix will make the interpretation more clear.

We can extend the change of variables to make the $\bar \sigma$ integral Gaussian to higher orders in $\lambda$. For example, at order $\lambda^2$ we have
\begin{align}
\bar\sigma_{\k}' =& \ \bar\sigma_{\k} + \frac{\lambda}{2E_{k} (k_1 + k_2 + E_{k})} (\bar\phi_1 \bar\phi_2 + \bar\phi_1' \bar\phi_2') \nonumber \\
&- \frac{2\lambda^2}{4E_k(k_1 + E_3 + k )(k_2 + E_k + k )(k_1 + E_3 + k_2 + E_k )}  (\bar\phi_1\bar\phi_2 + \bar\phi_1' \bar\phi_2') \bar\sigma_3 + ``1\leftrightarrow 2" \nonumber \\
&+ \cdots.
\end{align}
Note that the second line is of order $\lambda^2$, which introduces a correction to $\bar\sigma$ starting at order $\lambda^3$. For our purposes here it suffices to stop at order $\lambda^2$, so in the following we will ignore the $\bar\phi^2 \bar\sigma^2$ term, as well as the $\O(\lambda^3)$ terms in defining the new variable. 
The immediate consequence of the change of variable is (at leading order in $\lambda$)
\beq
\bar\sigma - \frac{\psi_{\phi^2\sigma}\bar\phi^2+\psi^*_{\phi^2\sigma}\bar\phi'^2}{\psi_{\sigma^2} + \psi_{\sigma^2}^* } =0 \ .
\eeq
After setting $\bar\phi = \bar\phi'$ and realizing that $\psi_{\phi^2\sigma}=\psi^*_{\phi^2\sigma}$ and $\psi_{\sigma^2}=\psi^*_{\sigma^2}$, this is
\beq
\bar\sigma + \frac{\lambda}{E_{k_3} (k_1 + k_2 + E_{k_3})} \bar\phi^2 =0 \ .
\eeq
It is then straightforward to see that in order to calculate $\langle \bar\phi_1 \bar\phi_2 \bar\sigma_3 \rangle$, one can replace $\bar\sigma$ with its biased value, such that 
\begin{equation}
    \langle \bar\phi_1 \bar\phi_2 \bar\sigma_3 \rangle \rightarrow \langle \bar\phi_1 \bar\phi_2 \frac{-\lambda}{(k_3 + k_4 + E_{\k_3+\k_4})E_{\vec k}} \bar\phi_3 \bar\phi_4 \rangle = - \frac{\lambda}{2k_1 k_2 E_3(k_{12}+E_3)} \ .
\end{equation}
This recovers exactly the desired result (\ref{eq:in-in_phi^2sigma}). Both the prime on $\bar\phi'$ and the complex conjugate on the wavefunction coefficients are statements that we really need to consider the density matrix instead of only the wavefunction. The evolution of the density matrix can be understood in the context of open-EFTs which have a number of cosmological applications~\cite{Burgess:2015ajz,Burgess:2022rdo,Salcedo:2024smn,Burgess:2024eng,Burgess:2024heo,Colas:2024ysu,Colas:2024lse}.

Another immediate result is that the new coefficient of the $\bar\phi^4$ term is now 
\begin{align}
    \log \rho[\bar \phi,\bar\phi] &\supset r_{\phi_1\phi_2\phi_3\phi_4} \bar\phi_1 \bar\phi_2 \bar\phi_3 \bar\phi_4 \nonumber \\
   & =  \frac{\lambda^2(E_k + k_1 + k_2 + k_3 + k_4)}{(k_1 + k_2 + E_k )(k_3 + k_4 + E_k )(k_1 + k_2 + k_3 + k_4 )E_k} \bar\phi_1 \bar\phi_2 \bar\phi_3 \bar\phi_4 \label{eq:psi_eft_phi4} \ .
\end{align}
Since there is no more cubic terms in the reduced density matrix, we have directly
\begin{align} 
    \langle \phi^4 \rangle &\supset 8\times \frac{r^{EFT}_{\phi_1\phi_2\phi_3\phi_4}}{r_{\phi_1^2} r_{\phi_2^2} r_{\phi_3^2} r_{\phi_4^2}} \nonumber \\
    &= \frac{\lambda^2(E_k + k_1 + k_2 + k_3 + k_4)}{ 2k_1 k_2 k_3 k_4 E_k (k_1 + k_2 + E_k )(k_3 + k_4 + E_k )(k_1 + k_2 + k_3 + k_4 )} \label{eq:expected_phi^4} \ .
\end{align}
This matches our previous $s$ channel calculations (\ref{eq:in-in_phi^4}) and (\ref{eq:phi^4_wavefunction_full}). One interesting observation is that the correction to $r_{\phi^4}$ coming from completing the square is 
\begin{equation}\label{eq:phi^4_from_square}
    \log \rho[\bar\phi,\bar\phi'] \supset + \frac{\lambda^2}{2(k_1 + k_2 + E_k)(k_3 + k_4 + E_k)E_k} (\bar\phi_{\vec k_1} \bar \phi_{\vec k_2}+\phi'_{\vec k_1} \bar \phi'_{\vec k_2})(\bar \phi_{\vec k_3}\bar \phi_{\vec k_4}+\bar \phi'_{\vec k_3}\bar \phi'_{\vec k_4} ) \ .
\end{equation}
The contribution of this term to the final result, after taking $\bar\phi = \bar\phi'$, is equal to that of the second term in (\ref{eq:phi^4_wavefunction_full}), which is a summation of terms shown in Fig. \ref{fig:tree_exchange_semilocal}. Since $\bar\sigma \neq 0$, their contributions are non-zero. Later in section \ref{sec:ERG}, we will derive how they appear in the exact RG formalism.

On the other hand, if we started with the effective action after integrating out the heavy field in the traditional sense, 
\begin{equation}
    S_{EFT} = \int d^4x  \left[ -\frac{1}{2}\partial_\mu\phi\partial^\mu\phi + \frac{\lambda^2}{2} \phi^2 \frac{1}{M^2 - \Box} \phi^2 \right] \ ,
\end{equation}
and use it to calculate the in-in correlators directly or through the wavefunction, we would not be able to recover the correct result for trispectrum. Most obviously, this action fails to capture terms with odd powers of $M$. As mentioned before, the odd part of the correction corresponds to semi-local terms. Specifically, Equation~(\ref{eq:missing_terms}) shows that the terms not captured by this effective action are analytic in two or more momenta, which correspond to $\delta$-function localized terms in position space. One may take the point of view that these are ultra-short distance effects that cannot be captured in the EFT (i.e.~we should only concern ourselves with correlators at separated points). However, it is possible that these can be captured by local corrections to the probability distribution function (or density matrix) which could be matched in the EFT. We will discuss this in detail in Section~\ref{sec:ERG}.

The even power in $1/M$ corrections, on the other hand, can be understood as the consequence of having a well-defined variational principle. Here we show, following mainly \cite{Bittermann:2022nfh, Dyer:2008hb}, how this works for the $1/M^4$ correction and discuss this topic in more detail in Appendix \ref{app:A}. In the EFT action, the $1/M^4$ term is of the form $\lambda^2 \phi^2 (\Box / 2M^4) \phi^2$. The spatial derivatives are harmless since we are integrating over the full spatial volume. However, the time derivative might be problematic, since we have a non-trivial boundary at $t=t_0$. The variation of such a term is 
\begin{equation}
    \delta S \supset \int d^4x \; \frac{\lambda}{2} \left( 2\phi \delta\phi \frac{-\partial_t^2}{M^4} \phi^2 +  2 \phi^2 \frac{-\partial_t^2}{M^4} \phi\delta\phi  \right) \ .
\end{equation}
Varying the first term with respect to $\delta \phi$ is straightforward, while for the second term we need to do some integration by parts, 
\begin{equation}
    \int d^4x \; \lambda^2 \phi^2 \frac{-\partial_t^2}{M^4} \phi\delta \phi = \int d^4x \; \lambda^2 (-\partial_t^2 \phi^2) \frac{1}{M^4} \phi \delta \phi - \lambda^2 \phi^2 \frac{\partial_t}{M^4} \phi \delta\phi \bigg|_{t_i}^{t_f} + \lambda^2 (\partial_t \phi^2) \frac{1}{M^4} \phi \delta\phi \bigg|_{t_i}^{t_f} \ .
\end{equation}
At this order the equation of motion is a second order differential equation with respect to time. Therefore, we are free to choose two boundary conditions. Setting $\delta \phi (t_i) = \delta \phi(t_f) = 0 $ uses up the two boundary conditions, and further setting $\partial_t\delta \phi (t_i) = \partial_t\delta \phi(t_f) = 0 $ will over-constrain the system. As a result, to have a well-defined variational problem, one can choose to set $\delta \phi (t_i) = \delta \phi(t_f) = 0 $ and add a boundary term $\lambda^2 \phi^2 \frac{\partial_t}{M^4} \phi \delta\phi \big|_{t_i}^{t_f}$ to cancel out the extra boundary term. That is, for the action to have a well defined variational problem, the $1/M^4$ term should be 
\begin{equation}
    S_{int} \supset \int d^4x \; \lambda^2 \phi^2 \frac{\Box}{2M^4} \phi^2 + \int d^3x \; \lambda^2 \bar\phi^2 \frac{\partial_t}{2M^4} \bar\phi^2  \ .
\end{equation}
Carrying out the standard procedure in the in-in calculation, this translates to the terms
\begin{equation}
    - \frac{\lambda^2}{k_T M^4}\frac{2\times 2}{16 k_1 k_2 k_3 k_4} (k_{12}^2+k_{34}^2 - 2 k_5^2) +  \frac{\lambda^2}{M^4} \frac{2\times 2}{16 k_1 k_2 k_3 k_4}(k_{12}+k_{34}) = -\lambda^2 \frac{k_{12}k_{34}+k_5^2 }{2k_1 k_2 k_3 k_4 k_T M^4} \ .
\end{equation}
This agrees with the expansion in (\ref{eq:phi4_real_expansion}). 

One can use this procedure to reproduce all the corrections that are even in $1/M$ order by order, at least in principle. However, in practice, this procedure already gets complicated at the $1/M^6$ order. Plus, it does not tell us about the terms odd in $1/M$. Fortunately, one can generate the full correction more efficiently by being more careful about the EFT procedure. The key point is that we use different boundary conditions for the in-out and in-in calculations, so the classical solutions to be plugged back into the action are different. Here we adapt the argument given in Section 4 of \cite{Salcedo:2022aal}. For an in-out calculation the homogeneous part of the heavy field simply vanishes in the EFT limit, while for an in-in calculation, the inhomogeneous solution to the equation of motion of the heavy field is corrected by the homogeneous part at the boundary to match the boundary condition. More explicitly, the full solution to the equation of motion of $\sigma$ can be written as
\begin{equation}
    \sigma(\vec k,t) = C_1 e^{iE_k t} + C_2 e^{-iE_kt} - \frac{\lambda}{M^2 - \Box}[\phi^2](\vec k,t) \ ,
\end{equation}
where $C_1$ and $C_2$ may depend on $\vec k$ only, and $[\phi^2](\k,t)$ is the fourier transform of $\phi^2(\x,t)$. In the usual in-out calculation, the boundary conditions are such that we do not see propagating heavy degrees of freedom in the asymptotic states. That sets $C_1 = C_2 = 0$, and we only plug the inhomogeneous solution back into the action for the tree level EFT. For in-in, however, the boundary conditions are now $\sigma(t\rightarrow-\infty)=0$ and $\sigma(t = t_0 = 0) = \bar\sigma(\vec x)$, where $\bar\sigma$ can be any real number. Apparently we need a different choice of $C_1$ and $C_2$. With our $i\epsilon$ prescription, the condition that $\sigma(t\rightarrow-\infty_{-})=0$, where $\infty_\pm = \infty(1\pm i\epsilon)$, tells us that $C_2 = 0$. To satisfy $\sigma(t = 0) = \bar\sigma(\vec x)$, we choose
\begin{equation}
    C_1 = \bar\sigma + \frac{\lambda}{M^2 - \Box}\bar\phi^2 \ ,
\end{equation}
where the second term cancels the inhomogeneous solution at $t=0$, and the value of $\sigma$ at the boundary is simply $\bar\sigma$. We are now left to choose the proper $\bar\sigma$ for tree level EFT. Turns out, we have already done the calculation to choose $\bar\sigma$; that is,
\beq
\bar\sigma_k - \frac{\psi_{\phi^2\sigma}+\psi^*_{\phi^2\sigma}}{\psi_{\sigma^2} + \psi_{\sigma^2}^* }\bar\phi^2 = \bar\sigma_k + \frac{\lambda}{E_k (k_1 + k_2 + E_k)} \bar\phi_1\bar\phi_2 =0 \ .
\eeq
It is then straightforward to plug everything back and calculate the tree level action. The full solution to $\sigma$, to be plugged into the action, is
\begin{align}\label{eq:sigma_wavefunctional_sol}
    \sigma(\k,t) =& e^{iE_k t} \left(\frac{\lambda}{-(k_1+k_2+ E_k)E_k} + \frac{\lambda}{M^2 + k^2 - (k_1+k_2)^2} \right)\bar \phi_1 \bar\phi_2 \nonumber \\
    &- \frac{\lambda}{M^2 + k^2 - (k_1+k_2)^2}\phi_1(t)\phi_2(t) \ .
\end{align}
The momenta involved satisfy $\k = \k_1 + \k_2$. The interaction part of the effective action, after plugging (\ref{eq:sigma_wavefunctional_sol}) back into the original action, is
\begin{equation}\label{eq:in-in_S_EFT}
    S \supset \frac{1}{2}\int d^4x  \; \lambda^2 \phi^2 \frac{1}{M^2 - \Box} \phi^2 + \frac{i}{2} \int d^3x \lambda^2\left( \frac{1}{M^2 - \Box} + \frac{1}{(i\partial_t - E)E}  \right)\bar\phi^2 \frac{i\partial_t + E}{M^2 - \Box} \bar\phi^2 \ ,
\end{equation}
where we have used the equation of motion and integration by parts to rearrange the terms, and we use $\partial_t \bar\phi_k$ to represent $\lim_{t\rightarrow 0} \partial_t \phi_k(t) = ik \bar\phi_k$.\footnote{This is the relation on the time-ordered path. On the anti-time-ordered path we have $\lim_{t\rightarrow 0} \partial_t \phi_k(t) = -ik \bar\phi_k$. At the same time, the signs in front of explicit factors of $i$'s should be flipped, resulting in a term that appears to be non-Hermitian at face value. See (\ref{eq:M^5_expansion_anticommutator}) as an example.}  In terms of wavefunctional coefficients, this gives, in the s channel,
\begin{align}
   \psi^{EFT}_{\phi^4} \supset& \frac{\lambda^2}{4}\left( \frac{2E^2 - \omega_{12}^2 - \omega_{34}^2}{\omega_T(E^2 - \omega_{12}^2)(E^2 - \omega_{34}^2)} - \frac{2E - \omega_T}{(E^2 - \omega_{12}^2)(E^2 - \omega_{34}^2)} + \frac{2}{(E+\omega_{12})(E+\omega_{34})E} \right) \nonumber \\
   =& \frac{\lambda^2(E + \omega_T)}{2(\omega_{12} + E )(\omega_{34} + E )\omega_TE} \ .
\end{align}
The $t$ and $u$ channels are permutations of the same result. We have symmetrized between $12 \leftrightarrow 34$, and omitted the subscript on $E$ as its value is obvious. This is exactly (\ref{eq:psi_eft_phi4}), if we take $\omega_{ij} = k_i + k_j$, and $\omega_T = k_1 + k_2 + k_3 + k_4$. The first term is the contribution from the traditional EFT action, and it obviously contains terms that are only even power in $M$, since $E^2 = M^2 + k^2$. The odd powers of $M$ come from the boundary corrections in the last two terms. A closer look reveals that the third term is the equivalent of the correction from completing the square in the wavefunction, or from the boundary contributions of the exchange diagrams. The second term is there to cancel out the boundary contributions from the traditional EFT term. Both the second and the third terms contain both even and odd power in $1/M$ terms. 

If one were to set $\bar\sigma_k = 0$, one would obtain a result without the last term, or, 
\begin{align}
   \tilde\psi_{\phi^4} \supset& \frac{\lambda^2}{4}\left( \frac{2E^2 - \omega_{12}^2 - \omega_{34}^2}{\omega_T(E^2 - \omega_{12}^2)(E^2 - \omega_{34}^2)} - \frac{2E - \omega_T}{(E^2 - \omega_{12}^2)(E^2 - \omega_{34}^2)} \right)\\
   =& \frac{\lambda^2}{2(\omega_{12} + E )(\omega_{34} + E )\omega_T} \ .
\end{align}
This is in fact the original $\psi_{\phi^4}$ in the UV wavefunction. This is expected since setting $\bar\sigma_k = 0$ implies that there should not be any terms with the heavy field line touching the boundary. In other words, the contributions of the terms in Fig. \ref{fig:tree_exchange_semilocal} vanish. This also agrees with the result when we simply set all $\bar\sigma$ in the wavefunction to zero. We see that the change of variable for $\bar\sigma$, the contribution from diagrams with heavy field line touching the boundary, and the non-zero value of $\bar\sigma_k$ in the solution are different manifestations of the same statement.

One can also expand the boundary correction terms in powers of $1/M$, 
\begin{align}
    S \supset & \frac{i}{2} \int d^3x \lambda^2\left( \frac{1}{M^2 - \Box} + \frac{1}{(i\partial_t - E)E}  \right)\bar\phi^2 \frac{i\partial_t + E}{M^2 - \Box} \bar\phi^2  \\
    =& \int d^3x\; \lambda^2 \left( \frac{\bar\phi^2 \partial_t \bar\phi^2}{2M^4} + i \frac{\partial_t \bar\phi^2 \partial_t \bar\phi^2}{2M^5} - \frac{1}{2M^6} \left( \bar\phi^2 \partial_t^3 \bar\phi^2 + \partial_t \phi^2 \partial_t^2 \phi^2 + 2 \partial_t \phi^2 \vec\nabla^2 \phi^2 \right) + \O(\frac{1}{M^7}) \right) \ . \nonumber 
\end{align}
We see explicitly the even and odd power in $1/M$ corrections. Carefully including these boundary corrections for both the time-ordered and anti-time-ordered part in the in-in calculation, the corrections coming from the $1/M^5$ term gives the following contribution to the trispectrum, 
\begin{align} \label{eq:M^5_expansion_anticommutator}
    \langle \bar\phi_1 \bar\phi_2 \bar\phi_3 \bar\phi_4 \rangle \supset& - \int d^3 x \; \lambda^2 \langle \frac{\partial_t \bar\phi^2 \partial_t \bar\phi^2}{2M^5} \bar\phi_1 \bar\phi_2 \bar\phi_3 \bar\phi_4 + \bar\phi_1 \bar\phi_2 \bar\phi_3 \bar\phi_4 \frac{\partial_t \bar\phi^2 \partial_t \bar\phi^2}{2M^5} \rangle \nonumber \\
    =& \frac{\lambda^2}{2 k_1 k_2 k_3 k_4 M^5} (k_{12}k_{34}+k_{13}k_{24}+k_{14}k_{23})(2\pi)^3\delta^3(\k_1 +\k_2 +\k_3 +\k_4) \ .
\end{align}
This is exactly the expected $1/M^5$ contribution, according to (\ref{eq:in-in_phi^4}) and (\ref{eq:phi4_real_expansion}). Later we will see how this same term arises in the exact RG formalism\footnote{To compare to equation (\ref{eq:semilocal_ERG}), note that here in (\ref{eq:M^5_expansion_anticommutator}) the $\lambda$ has mass dimension one, so the overall mass dimension coming from $\lambda^2/M^5$ is $-3$. In (\ref{eq:semilocal_ERG}), $g$ is dimensionless.}. However, this appears as an anti-commutator for a scalar field at the boundary. Nevertheless, as mentioned before, we are saved by the fact that all of the odd power in $1/M$ terms are semi-local. Therefore, this anti-commutator term is not expected to be captured by conventional EFT. On the other hand, all the even power terms contribute through a commutator, and they can be understood as local field redefinitions on the boundary. 

In \cite{Burgess:2024eng, Burgess:2024heo}, it is argued that deviations from a naive pure state result can be eliminated with a certain choice of the order of the limits $\varepsilon \rightarrow 0$ and $M\rightarrow \infty$. Here $i\varepsilon$ is a choice of a specific UV regulator to make the momentum integral converge at high $k$, $\varepsilon \sim 1/\Lambda$, and is a different regulator from our $\epsilon$ above. It is defined to be the small imaginary part of the time difference in the propagator, $\varepsilon = -\text{Im} (t-t_0)$, and is usually held constant prior to the momentum integral. Taking $\varepsilon \rightarrow 0$ means that we have an infinite resolution for UV effects. It is then not very surprising that we see more than usual EFT effects. One way to decide which order of limit to take is to compare the mass of the heavy field $M$ to the cutoff energy $\Lambda$. In the case $M \gg \Lambda$, one should first expand in $M \rightarrow \infty$. When $M \ll \Lambda $, one should first take $\varepsilon \rightarrow 0$.

In our example, expanding in $1/M$, all propagators that involve $\sigma$ are exponentially suppressed by $\sim e^{-\varepsilon M}$, except for terms that are local in time. This leads to exactly the EFT term in the bulk, plus corrections on the boundary, both local in time. The fact that we see precisely the usual EFT term in the bulk implies that we can treat the system as evolving with only the effective action up to exponentially suppressed terms in the bulk. This corresponds to the claim that if we first expand in $M \rightarrow \infty$, we will kill the time derivative of the purity, up to exponentially small terms. However, polynomial corrections can still exist on the boundary, and we have seen explicitly that the state we end up with is not a pure state. The way to understand this is that indeed the purity does not evolve in the bulk, as a consequence of taking the initial time to $-\infty(1-i\epsilon)$ (see section 3.1 in \cite{Burgess:2024eng}). However, the system maintains a purity smaller than one all the time, reflected in the correction terms and the mixed state at the boundary.

\section{Exact Results in Flat Space}\label{sec:ERG}

Our most complete understanding of EFT and RG comes from the non-perturbative RG defined by Polchinski~\cite{Polchinski:1983gv}. This approach provides, in a particular scheme, exactly how the action must change after integrating out a shell of momentum. Importantly, we will see that for in-in correlators, the non-perturbative RG includes both the standard change to the effective action and local boundary terms.

\subsection{Exact Wilsonian RG}

In the context of inflation, we are used to evaluating correlation functions in the interacting vacuum (in-in correlators). We can define an in-in correlator in perturbation theory in a variety of spacetime backgrounds as~\cite{Weinberg:2005vy}
\beq
\langle {\cal O}(t)\rangle=\left\langle\left[\bar{\text{T}} \exp \left(i \int_{-\infty(1+i \epsilon)}^t \d t\, H_{\rm int}(t) \right)\right] {\cal O}^{\rm int}(t)\left[\text{T} \exp \left(-i \int_{-\infty(1-i \epsilon)}^t \d t\, H_{\rm int}(t)\right)\right]\right\rangle \ ,
\eeq
where ${\cal O}(t)$ is an operator defined at a single time $t$, usually in terms of products of local fields, and ${\cal O}^{\rm int}(t)$ is the same operator defined in terms of interaction picture fields. The $i\epsilon$ prescription is important for defining the interacting vacuum of the theory. 

We can also calculate in-in correlators from the wavefunction of the universe. Given a real scalar field $\varphi(\x,t)$ whose equal time correlators we wish to calculate at a time $t_0$, we can write
\beq
\langle {\cal O}(t)\rangle = \int D{\bar \varphi}(\x) \O[{\bar \varphi}(\x)] |\Psi[{\bar \varphi}(\x);t]|^2 \ ,
\eeq
where we defined $\varphi(\x,t_0) \equiv \bar \varphi(\x)$. In flat space, we can write the wavefunction in terms of the path integral
\beq
\Psi[{\bar \varphi};t] = \int^{{\bar \varphi}} D\varphi(\x,t) \exp\left( i \int^{t_0}_{-\infty(1-i \epsilon)} {\cal L}[\varphi(\x,t)] \right) \ ,
\eeq
where
\beq
{\cal L} =  \int \frac{d^3 k}{(2\pi)^3} \frac{K^{-1}(k^2/\Lambda_0^2)}{2}(\dot \varphi^2 -\partial_i \varphi \partial^i \varphi) + {\cal L}_{\rm int}[\varphi] \ , 
\eeq
and $K(z)$ is a regulator\footnote{Note that in this section $K$ is not the bulk to boundary propagator.} that is defined so that $K(z\ll 1) \to 0$ and $K(z\gg 1) \to 1$. In this regard, we can recognize the perturbative formula as the expansion of the wavefunction expression around the wavefunction of the free theory.

In order to write a general expression for the RG flow of the in-in correlator, we define a partition function
\beq
Z[J] = \int D{\bar \varphi} \exp \left(  \int \frac{d^3 k}{(2\pi)^3} \, J(-\k) {\bar \varphi}(\k)    \right) |\Psi[{\bar \varphi}; t_0] |^2 \ .
\eeq
Any in-in correlator of $\O[{\bar \varphi}]$ can be derived from this partition function by functional derivatives with respect to $J(-\k)$. We can write this as time-ordered path integral 
\beq
Z[J] = \int_\gamma D\varphi \; T \exp \left(  i S_{\gamma} [\varphi] +\int \frac{d^3 k}{(2\pi)^3}J(-\k) \bar \varphi(\k)  \right) \ ,
\eeq
where $S_{\gamma} = \int_\gamma dt {\cal L} + S_{\partial}$ is the action, with $S_\partial$ being a boundary-term needed to make the variational principle well defined, and $\gamma$ is the contour $\gamma:(-\infty(1-i\epsilon),t_0] \cup [t_0,-\infty(1+i\epsilon))$. Here $T$ denotes time-ordering along the contour. 

Now we want to integrate out a shell of momentum from $k \in [\Lambda_0-\delta \Lambda, \Lambda_0]$ so that we can lower the cutoff to $\Lambda =\Lambda_0-\delta \Lambda$, and then repeat to arrive at a theory with $\Lambda \ll \Lambda_0$. After performing the path integral, we will arrive at a partition function
\begin{align}
Z[J,\Lambda] =& \int_\gamma D\varphi \, T \exp \bigg( i S_0[\varphi(\x,t), \Lambda]  + i S_I[\varphi(\x,t), \Lambda]  +\int \frac{d^3 k}{(2\pi)^3}J(-\k) \bar \varphi(\k) \bigg) \ , \\
S_0[\varphi(\x,t), \Lambda]  \equiv & \int_\gamma dt \int \frac{d^3 k}{(2\pi)^3} \frac{K^{-1}(k^2/\Lambda^2)}{2}(\dot \varphi^2 -\partial_i \varphi \partial^i \varphi)   \ , \end{align}
where we defined the interaction term in the action after integrating out as
\beq
S_I =  S_{\rm int,\Lambda} + S_{\partial,\Lambda} \ .
\eeq
The bulk action, $S_{\rm int,\Lambda} = \int_\gamma {\cal L}_{\rm int}$, is the effective interaction derived from the local evolution, and $S_{\partial,\Lambda}$ is the full boundary term localized at $t_0$. It is important that $ S_{\partial,\Lambda}$ contain operators that do not commute with ${\bar \varphi}$ and therefore the precise definition of this term will be important. As we saw above, this partition function is not generated by a wavefunction, but is now described by a reduced density matrix
\beq\label{eq:ZLJ}
Z[J,\Lambda] = \int D{\bar \varphi} \exp \left(  \int \frac{d^3 k}{(2\pi)^3}  s_\partial({\bar \varphi},J,t_0) \right) \rho_\Lambda[{\bar \varphi},{\bar \varphi}; t_0] \ .
\eeq
The boundary action, $S_\partial$, defined in the partition function mixes the contributions to the density matrix, $\rho$, and the operators and sources $s_\partial({\bar \varphi},J,t_0)$. This distinction is useful to separate contributions from field redefinition, ${\bar \varphi} \to F[{\bar \varphi}]$, and the change to the density matrix in a fixed basis.

The partition function $Z[J,\Lambda]$ is necessarily independent of $\Lambda$ so that $\frac{d}{d\Lambda} Z[J,\Lambda] =0$. Specifically, all we did in defining $Z[J,\Lambda]$ was to perform part of the integral present in $Z[J]$. Therefore, differentiating the expression we have 
\begin{equation}
    \frac{d}{d\log \Lambda} Z[\Lambda,J]  = \int \mathcal{D}\phi \; i \frac{d}{d\log \Lambda} T \left( S_0 + S_{I}  \right) e^{iS_0 +i S_I + \int \frac{d^3k}{(2\pi)^3} J(-\k)\bar\varphi(\k) } \ .
\end{equation}
It is natural to define the evolution of $S_I$ under RG so that it mimics the non-perturbative RG of the Euclidean particle function as much as possible. We should find that the effective action should be the same as what we would find for in-out correlators. In-in correlators contain the S-matrix as the residue of the total energy pole and therefore we should find that these residues are calculable using the scattering amplitude calculated within a conventional EFT. Given this expectation, the most natural resolution is to change $S_\partial$ such that the integrand is the same total derivative as we find the Euclidean and/or in-out partition function.

There is a straightforward method to realize this behavior, following~\cite{Goldman:2024cvx}. First, we note that 
\beq
\frac{\partial}{\partial \log \Lambda} S_0 = - \frac{1}{2} \int \frac{d^3 k}{(2\pi)^3} \frac{dK(\tfrac{k^2}{\Lambda^2})}{d\log \Lambda}   \int_\gamma dt \int_\gamma dt' \left( \frac{\delta S_0}{\delta \varphi(\k,t)} \right)  \, G(t,t', \k) \,\left( \frac{\delta S_0}{\delta \varphi(-\k,t')}\right) + \delta S_{\partial,2} \ ,
\eeq
where $G(t,t',\k)$ is the path-ordered Green's function that obeys,
\beq
\left(\frac{\partial^2}{\partial t^2 } + k^2 \right)G(t,t',\k) = \delta(t-t') \ , 
\eeq
 and $\delta S_{\partial, 2}$ is a quadratic boundary term derived in Appendix~\ref{app:B}. By virtue of the Schwinger-Dyson equations,
\begin{equation}
   \mathcal{O} \frac{\delta S_0}{\delta \phi}  =  \frac{\delta \mathcal{O}}{\delta \phi} - \mathcal{O} \frac{\delta S_{I}}{\delta \phi}   \ ,
\end{equation}
we can exchange $S_0 \to S_I$ to find
\begin{align}
\frac{d}{d\log \Lambda} S_0 =& \int\frac{d^3 k}{(2\pi)^3} \int_\gamma dt \int_\gamma dt'\frac{d}{d\log \Lambda} K \Bigg(-\frac{1}{2} \frac{\delta  S_I}{\delta \varphi(\k,t)}  \, G(t,t', \k) \, \frac{\delta  S_I}{\delta \varphi(-\k,t')} \\
&+ \frac{1}{2} \frac{\delta^2 S_I}{\delta \varphi(\k,t) \delta \varphi(-\k,t')}  G(t,t',\k) \Bigg) + \delta S_{\partial,2}   \ .
\end{align}
Therefore, if 
\beq \label{eq:ERG_S_I_equation}
\frac{d}{d\log \Lambda} S_I = \int\frac{d^3 k}{(2\pi)^3} \int_\gamma dt dt'\left(\frac{1}{2} \frac{\delta S_I}{\delta \varphi(t)} G(t,t') \frac{\delta S_I}{\delta \varphi(t')} -\frac{1}{2} \left(\frac{\delta^2 S_I}{\delta \varphi(t) \delta \varphi(t')}  G(t,t')\right) \right) - \delta S_{\partial,2} \ , 
\eeq
we will have $\partial_{\log \Lambda} Z[J,\Lambda] = 0$. Up to the quadratic boundary terms (which are unimportant to this discussion), this expression looks identical to the equivalent formula for in-out or Euclidean partition functions. However, as we have already seen in an explicit example, it is not the case that integrating out leads to a local action (or a pure state).

We would like to understand in what sense the change to $S_I$ is local. We suppose that $S_I$ is local in the sense that 
\beq
S_I = \int_{-\infty(1-i \epsilon)}^{t_0} dt {\cal L}_{\rm int}[\varphi(t)] -\int_{-\infty(1+i \epsilon)}^{t_0} dt' {\cal L}_{\rm int}[\varphi(t)'] + {\cal Q}[\varphi,t_0] \ ,
\eeq
where ${\cal Q}$ is an operator define in terms of $\varphi$ and its derivatives at $t_0$. If $S_I$ local, the second term in the expression (\ref{eq:ERG_S_I_equation}) is always local as well,
\beq
-\frac{1}{2} \frac{\delta^2 S_I}{\delta \varphi(t) \delta \varphi(t')}  G(t,t') = - \frac{\delta^2 \int dt'' {\cal L}_{\rm int}(t'')}{\delta \varphi(t) \delta \varphi(t')} G(t,t') \propto  \delta(t-t') G(t,t) \ .
\eeq
In contrast, locality of the first term depends on the properties of the Green's function. If both times are in the same part of the contour, $t,t' \in (-\infty(1 \pm i\epsilon),t_0)$, then the Green's function is exponentially suppressed in $|t-t'|$,
\beq
G(t,t')= \frac{1}{2k} \exp(\pm ik |t-t'|(1\pm i \epsilon)) \ .
\eeq
Therefore, if we define
\beq
\frac{\delta S_I}{\delta \varphi(t)} = \delta {\cal L}(t) \ ,
\eeq
then 
\bea
\int^{t_0}_{-\infty(1\pm i \epsilon) } dt dt' \frac{1}{2} \frac{\delta S_I}{\delta \varphi(t)} G(t,t') \frac{\delta S_I}{\delta \varphi(t')} &=& \int^{t_0}_{-\infty(1\pm i \epsilon) } dt \int d
(\Delta t) \, \delta {\cal L}(t) \delta {\cal L}(t+\Delta t)G(\Delta t) \\
&=& \int^{t_0}_{-\infty(1\pm i \epsilon) } dt \sum_j {\cal Q}_j[\varphi,t_0] \ .
\eea
In the last step, we used the fact that the integral over $\Delta t = t-t'$ is exponentially suppressed in $\Delta t$ to write $\delta {\cal L}(t+\Delta t)$ as an expansion in local operators defined at $t_0$, ${\cal Q}_j$, multiplied by a convergent integral over $\Delta t$. Furthermore, since both terms are on the same side of the path from $t_0$, there is no issue with time ordering with respect to the sources $J {\bar \varphi}$.

Now we must consider the remaining terms that are not manifestly local and do not arise in the usual in-out expressions. To be concrete, we take $t \in (-\infty(1+i\epsilon),t_0]$ and $t' \in (-\infty(1-i\epsilon),t_0]$ so that 
\beq
\frac{d}{d\log \Lambda} S_I  \supset \int^{t_0}_{-\infty(1+ i \epsilon) } dt \int^{t_0}_{-\infty(1- i \epsilon) } dt' \frac{1}{2} \frac{\delta S_I}{\delta \varphi(t)} G(t,t') \frac{\delta S_I}{\delta \varphi(t')} \ .
\eeq
Because the two integrals have different $i\epsilon$ prescriptions, there is no reason they should be expressed as a local term in the action integrated over the contour. Such a term would be non-local along the contour and thus would not be related to a local action in Euclidean signature. However, the Green's function is also no longer exponentially suppressed with respect to $|t-t'|$ but instead takes the form
\beq
G(t,t',\k) = \frac{1}{2k} \exp(ik |t-t_0|(1+ i \epsilon) -ik |t'-t_0|(1- i \epsilon) ) \ .
\eeq
We see now that both terms are exponentially suppressed in $|t-t_0|$. As a result, we can expand both terms in $t= t_0- \Delta t$ and $t'=t_0 - \Delta t$ to find a correction to the boundary term, $S_{\partial,{\rm int}}$
\beq
\frac{d}{d\log \Lambda} S_{\partial,{\rm int}}  \supset  \int^{t_0}_{-\infty(1+ i \epsilon) } dt \int^{t_0}_{-\infty(1- i \epsilon) } dt' \frac{1}{2} \frac{\delta S_I}{\delta \varphi(t)} G(t,t') \frac{\delta S_I}{\delta \varphi(t')} \to \sum_{i,j} {\cal Q}_i(t_{0,+}) {\cal Q}_j(t_{0,-}) \ ,
\eeq
where ${\cal Q}_i(t_{0,+}) {\cal Q}_j(t_{0,-})$ the product of local operators defined at $t_{0,\pm}$ respectively. However, we cannot express this directly as a local operator at $t_0$. Specifically, if we have a term of the form
\beq\label{eq:exact_boundary_eg}
 S_{\partial,{\rm int}} \supset g \dot \varphi(t_{0,+}) \varphi(t_{0,-})^n \ ,
\eeq
then since $\dot \varphi$ does not commute with ${\bar \varphi}$, we have 
\beq
T( g \dot \varphi(t_{0,+}) \varphi(t_{0,-})^n J{\bar \varphi}) = T(g J {\bar \varphi} \dot \varphi(t_{0,-}) \varphi(t_{0,-})^n) + i J g {\bar \varphi}^{n} \ . 
\eeq
 Notice that the second term is equivalent to a field redefinition of ${\bar \varphi}$, $\bar \varphi \to \bar \varphi+ g\bar \varphi^n$. Moreover, the correlations of ${\bar \varphi}$ and $\dot \varphi$ are $\delta$-functions. As a result, the additional local terms involving $\dot \varphi$ will generate semi-local terms and field redefinitions. 
These are precisely the terms we found from integrating out the massive field in Section~\ref{sec:mass_flat}. 

As a concrete illustration, by integrating out, we will generate terms of the form
\beq
 S_{\partial,{\rm int}} \supset \int d^3 x \frac{g}{\Lambda^3} \dot \varphi(t_+) {\bar \varphi} \dot \varphi(t_-) {\bar \varphi} \ ,
\eeq
where $g$ is a dimensionless parameter and $\Lambda$ is our UV scale. Due to the time ordering, the contribution to the in-in four point correlator is non-local,
\bea \label{eq:semilocal_ERG}
\langle in|  \varphi(\k_1)\varphi(\k_2) \varphi(\k_3) \varphi(\k_4)  |in \rangle' &\supset& \frac{g}{\Lambda^3}\int d^3 x \langle \dot \varphi(t_+) \varphi(t_0)  \varphi(\k_1)\varphi(\k_2) \varphi(\k_3) \varphi(\k_4) \dot \varphi(t_-) \varphi(t_0) \rangle' \nonumber \\
&=& \frac{g}{\Lambda^3} \frac{1}{8 k_2 k_4} + {\rm permutations} \ .
\eea
Notice two important aspects of this correction: (1) by dimensional analysis it scales like an odd power of $\Lambda$ and (2) the corrections are semi-local terms (e.g. the first term is analytic in $k_1$ and $k_3$). This is a general outcome that results from the structure of these non-local ``boundary" terms.

The diagram relevant to the exact RG we found here is precisely the same term we found from integrating-out the massive field. We are integrating out a field that has an exchange interaction between the two pieces of the in-in contour. These terms are still written as an expansion in time and space derivatives. Because these do not commute with $J{\bar \varphi}$, they are not equivalent to a local change to the action, nor a local term in the wavefunction. In this precise sense, we need to treat the EFT as a density matrix if we wish to reproduce the exact form of the correlators. Yet, we also see that these terms are, at most, equivalent to semi-local terms in the correlator. One may alternatively take the point of view that local and semi-local terms are not calculable within the EFT (like contact terms) and thus need to be determined by matching. 

\subsection{Wavefunction Sum Rules}

Sum rules (dispersion relations) applied to scattering amplitudes are efficient encapsulation of EFT philosophy and decoupling~\cite{Donoghue:1996kw}. For example, additional subtractions remove sensitivity of the scattering amplitude to information from the UV, at the cost of additional local terms. This offers a precise example of how UV physics is encoded in the local interactions of the low energy theory. Furthermore, positivity bounds~\cite{Pham:1985cr,Adams:2006sv,deRham:2022hpx} have shown the power counting expectations based on EFT and dimensional analysis are actually required by causality and unitarity~\cite{Baumgart:2022yty,Distler:2006if,Manohar:2008tc,Arkani-Hamed:2020blm,Bellazzini:2020cot,Tolley:2020gtv,deRham:2021bll,Caron-Huot:2022ugt}.

It is extremely desirable to have the same level of understanding of the internal consistency of EFT on cosmological backgrounds. Although cosmology currently lacks an observable as robust as the S-matrix, progress has been made in several directions. One such approach is to study sum rules at the level of individual wavefunction coefficients~\cite{Salcedo:2022aal}. These share some analytic properties with the S-matrix and contain, at tree level, the S-matrix as the residues of the total energy poles.

Defining an off-shell wavefunction, the analytic structure of the wavefunction coefficients $\psi_n$ allow one to write a dispersion relation~\cite{Salcedo:2022aal} 
\begin{equation}
\left.\omega_T \psi_n(\{\omega\},\{\mathbf{k}\})\right|_{\omega_1=\omega_1^{\prime}}=\int_{-\infty}^0 \frac{d \omega_1}{2 \pi i} \frac{\operatorname{disc}\left(\omega_T \psi(\{\omega\},\{\mathbf{k}\})\right)}{\omega_1-\omega_1^{\prime}}+\operatorname{Res}_{\omega_1=\infty}\left(\frac{\omega_T \psi_n(\{\omega\},\{\mathbf{k}\})}{\omega_1-\omega_1^{\prime}}\right) \ ,
\end{equation}
where $\omega_i$ are the off-shell extensions of the energies (usually subject to $\omega_i =|\k_i|$ for a massless field on-shell). At face value, this expression also allows one to relate subtractions to additional contact terms. Concretely, given an analytic function $f(z)$ for $z$ off the real axis, we can write the dispersion relation
\begin{equation}
\frac{f\left(z_0\right) - f(0)}{z_0}=\frac{1}{2 \pi i} \oint_C \frac{1}{z}\frac{f(z)-f(0)}{z-z_0}  \to f(z)= f(0) + \frac{z_0}{2 \pi i} \oint_C \frac{1}{z}\frac{f(z)-f(0)}{z-z_0}\ .
\end{equation}
We see that we can suppress the integral at large $z$ at the cost of introducing a constant $f(0)$. Applied to the wavefunction coefficients, we get
\begin{equation}
\left.\omega_T \psi_n(\{\omega\},\{\mathbf{k}\})\right|_{\omega_1=\omega_1^{\prime}}=\left.\omega_T \psi_n(\{\omega\},\{\mathbf{k}\})\right|_{\omega_1=0} + \omega_1' \int_{-\infty}^0 \frac{d \omega_1}{2 \pi i} \frac{1}{\omega_1} \frac{\omega_T (\psi_n(\omega_1)-\psi_n(0))}{\omega_1-\omega_1'} \ .
\end{equation}
This would validate the intuition that information at large $\omega$ is encoded in contact interactions.

The ideal use of this formula would be to compute $\psi_{n, {\rm EFT}}$ as $\omega_1' \ll \Lambda_{\rm UV}$ and use this expression to relate the IR wavefunction coefficients to unknown UV physics that defines $\psi_n$ at $\omega_1 \gg \Lambda_{\rm UV}$. While this is a straightforward produce for S-matrix elements, we have already seen the challenge of defining $\psi_{\rm EFT}$ and relating it to the wavefunction coefficients in the UV. 

If we consider the model with a heavy field, in~\cite{Salcedo:2022aal} the effective wavefunction is defined by $\Psi[{\bar \varphi},\bar \sigma=0]$. In other words, we are not considering the full wavefunction but only the the slice of the wavefunction where $\bar\sigma =0$. This is not sufficient to calculate an in-in correlator in the EFT, where one needs to integrate over $\bar \sigma$ rather than forcing it to take a specific value. Nevertheless, setting $\bar \sigma = 0$ is a valid procedure on its own, and can be a useful regime for constraining the structure of the wavefunction. The more non-trivial challenge is that $\omega_n$ does not have an unambiguous definition that can be matched between the UV and the IR. Concretely, under a field redefinition $\bar \sigma \to \bar \sigma +g {\bar \varphi}^2$ we shift the wavefunction coefficient
\beq
\psi_{\varphi^4} \to \psi_{\varphi^4} + g \psi_{\varphi^2 \sigma} \ .
\eeq
In this precise sense, the UV object that appears in this expression is not well defined within the EFT description. More over, the ``EFT" that defines the wavefunction is not the EFT one gets from integrating out the short wavelength modes (which is described by a density matrix). Nevertheless, we have seen that these issues are pertain to the local terms in the wavefunction and therefore non-local terms should be robust to these kinds of issues. This is indeed what was found in~\cite{Salcedo:2022aal}, where the addition of boundary terms to the wavefunction was sufficient to match the UV expression.

\section{EFT and de Sitter Correlators }\label{sec:dS}

Effective theories in de Sitter space, and during inflation, are subject to different constraints than their flat space counterparts. Energy is not conserved in an expanding background, and therefore the split between hard and soft modes is more nuanced than for scattering processes that are highly constrained by kinematics. Concretely, massive particles do not completely decouple~\cite{Chen:2009zp,Baumann:2011nk,Assassi:2012zq,Flauger:2016idt} but their direct production is exponentially (Boltzmann) suppressed at large mass~\cite{Noumi:2012vr,Arkani-Hamed:2015bza,Lee:2016vti}. On the other hand, the expansion of spacetime creates large density fluctuations from quantum fluctuations and therefore has a physical meaning without explicitly introducing a detector or observer. The expansion of the universe does redshift away heavy fields and thus their role should decouple in time, even if their impact does not decouple in the statistics of the metric fluctuations. In this section we check the decoupling of the heavy field with an explicit example.

\subsection{Mellin Space}
For our purpose, it is more convenient to work with the Mellin space representations of the mode functions in de Sitter spacetime. The Mellin representation has both computational and conceptual advantages for perturbative calculations in both AdS~\cite{Fitzpatrick:2011ia,Fitzpatrick:2012cg,Fitzpatrick:2012yx,Fitzpatrick:2014vua} and cosmological backgrounds~\cite{Sleight:2019mgd,Sleight:2019hfp,Premkumar:2021mlz,Qin:2022fbv,Qin:2023ejc,Qin:2023bjk,Cohen:2024anu,Green:2024fsz,Qin:2024gtr}.

For a generic scalar in de Sitter background, the mode function $h_k$ satisfies 
\begin{equation}
    h_k(\tau) '' - \frac{d-1}{\tau} h'_k(\tau) + \left(k^2 + \frac{m^2}{\tau^2 H^2}\right) h_k(\tau) = 0 \ ,
\end{equation}
where we use $'$ to represent the derivative with respect to the conformal time $\tau$. The solution is given by 
\begin{equation}
    h_k(\tau) = H\sqrt{\frac{\pi}{4}}e^{-\frac{i\pi}{4}(1+2\nu)} (-\tau)^{d/2} H_{\nu}^{(2)}(-k\tau) \ ,
\end{equation}
where $H_\nu^{(2)}$ is the Hankel function of the second kind, and 
\begin{equation}
    \nu = \sqrt{\frac{9}{4}-\frac{m^2}{H^2}} \ .
\end{equation}
For the principal series, we have $m^2/H^2 > 9/4$, and as a result, $\nu$ is purely imaginary. To avoid confusion, we define, 
\begin{equation}
    \nu = \bigg| \sqrt{\frac{9}{4}-\frac{m^2}{H^2}} \bigg| \ ,
\end{equation}
so that $\nu$ is always real and positive, and we simply replace $\nu$ with $i\nu$ for the principal series solutions. Manipulations with these Hankel functions are technically very challenging. To make progress, it is often helpful to express the Hankel functions in terms of their Mellin space representations~\cite{Watson:1944:TBF},
\begin{align}
    i\pi e^{\frac{i \pi \nu}{2} }H_{\nu}^{(1)}(z) =& \int_{c-i\infty}^{c+i\infty} \frac{ds}{2\pi i} \Gamma(s+\frac{\nu}{2}) \Gamma(s-\frac{\nu}{2})\left( -\frac{iz}{2} \right)^{-2s}  \ , \\
    -i\pi e^{-\frac{i \pi \nu}{2}}H_{\nu}^{(2)}(z) =& \int_{c-i\infty}^{c+i\infty} \frac{ds}{2\pi i} \Gamma(s+\frac{\nu}{2}) \Gamma(s-\frac{\nu}{2})\left( \frac{iz}{2} \right)^{-2s}  \ .
\end{align}
Using these, in Mellin space the mode function for the complimentary series field $\phi$ is given by 
\begin{equation}
    f_k(\tau) = \frac{i}{\pi} H \sqrt{\frac{\pi}{4}} e^{-\frac{i\pi}{4}} (-\tau)^{d/2} \int_{c-i\infty}^{c+i\infty} \frac{ds}{2\pi i} \Gamma(s - \frac{\nu}{2}) \Gamma(s + \frac{\nu}{2}) (-\frac{ik\tau}{2})^{-2s} \ ,
\end{equation}
and for the principal series $\sigma$ is 
\begin{equation}\label{eq:prin_m}
    g_k(\tau) = \frac{i}{\pi} H \sqrt{\frac{\pi}{4}} e^{-\frac{i\pi}{4}} (-\tau)^{d/2} \int_{c-i\infty}^{c+i\infty} \frac{ds}{2\pi i} \Gamma(s - \frac{i\nu}{2}) \Gamma(s + \frac{i\nu}{2}) (-\frac{ik\tau}{2})^{-2s} \ .
\end{equation}
These formulas can be understood as reproducing the series expansion of the Hankel functions using the residue formula on the poles of the $\Gamma$-functions. The advantage of the Mellin representation in de Sitter correlator calculations is that the argument of the Hankel function factorizes out so that integrals against it could be performed directly. 

\subsection{Trispectrum and EFT in Mellin Space}
Our goal is to consider the same two field model discussed in Section~\ref{sec:mass_flat}, but now in de Sitter space. For the application to cosmology, we will use the de Sitter metric in flat FRW slicing,
\beq
ds^2 = -dt^2 + a(t)^2 d^2 x  = a(\tau)^2 (-d\tau^2 + dx^2) \ , \qquad a(t) = e^{H t} =- \frac{1}{H \tau} \ ,
\eeq
where $\tau$ is the conformal time. In these coordinates, the action becomes
\begin{equation}
    S = \int d\tau d^3 x \; \frac{1}{H^4 \tau^4} \left[ -\frac{1}{2}\partial_\mu\phi\partial^\mu\phi - \frac{1}{2}\partial_\mu\sigma\partial^\mu\sigma - \frac{1}{2}M^2 \sigma^2 - \lambda \phi^2 \sigma \right] \ .
\end{equation}
From the action, we can derive the equation of motion for $\sigma$,  
\begin{equation}
    \sigma '' - \frac{d-1}{\tau} \sigma' + \left(k^2 + \frac{M^2}{\tau^2 H^2}\right) \sigma = -\frac{\lambda}{\tau^2 H^2} \phi^2 \ ,
\end{equation}
where $d\to 3$ is number of spatial dimension. Using the equations of motion to write $\sigma$ in terms of $\phi^2$, the effective interactions of our EFT become 
\beq 
    S_{I,EFT} = \int \frac{d\tau d^3 x}{H^4 \tau^4} \; \frac{\lambda^2}{2 H^2} \phi^2 \frac{1}{\tau^2 \partial_\tau^2 - \tau (d-1) \partial_\tau + \frac{M^2}{H^2} + \tau^2 k_5^2} \phi^2 \ . 
\eeq
Since $\sigma$ is a massive field, its mode function should vanish as $k\tau \rightarrow 0$. As a result, we expect the boundary contributions to the effective wavefunction coming from the exchange diagrams to vanish. To check if this is true, we compare the trispectrum $\langle \phi^4 \rangle$ calculated as exchange diagrams using the full UV action, with the same result obtained as a contact diagram using the effective action. We lay out a more detailed calculation in Appendix \ref{app:C}, and report the main results here. 

The full result of the trispectrum calculation in Mellin space is not the most illuminating result to look at, as it involves special functions that are opaque to human eyes. Instead, we compare the two calculations at some intermediate stage. To begin with, the in-in calculation of the trispectrum using the full action is basically 
\begin{align}
    & 2\text{Re} \left( -\lambda^2 \int_{-\infty}^{\tau_0} \frac{d\tau_2}{H^4\tau_2^4} \int_{-\infty}^{\tau_2} \frac{d\tau_1}{H^4\tau_1^4} \langle \phi^2 \sigma (\tau_1) \phi^2 \sigma(\tau_2) \phi^4(\tau_0) - \phi^2 \sigma (\tau_1) \phi^4(\tau_0) \phi^2 \sigma(\tau_2) \rangle \right)  \\
    =& 2\text{Re} \bigg( -\lambda^2 f_1^*(\tau_0)f_2^*(\tau_0)f_3^*(\tau_0)f_4^*(\tau_0) \int_{-\infty}^{\tau_0} \frac{d\tau_2}{H^4\tau_2^4} \int_{-\infty}^{\tau_2} \frac{d\tau_1}{H^4\tau_1^4} f_1(\tau_1)f_2(\tau_1)f_3(\tau_2)f_4(\tau_2)g_5(\tau_1)g_5^*(\tau_2) \nonumber \\
    &+\lambda^2 f_1^*(\tau_0)f_2^*(\tau_0)f_3(\tau_0)f_4(\tau_0) \int_{-\infty}^{\tau_0} \frac{d\tau_2}{H^4\tau_2^4} \int_{-\infty}^{\tau_2} \frac{d\tau_1}{H^4\tau_1^4} f_1(\tau_1)f_2(\tau_1)f_3^*(\tau_2)f_4^*(\tau_2)g_5(\tau_1)g_5^*(\tau_2)\bigg) \nonumber \ ,
\end{align}
where $g_k$ is the mode function of $\sigma$, which is given by the principle series solution in Equation~(\ref{eq:prin_m}). We show in Appendix \ref{app:C} that the second term is exponentially suppressed in $M$, so we do not expect it to show up in the EFT. Therefore, we focus on the first term. Plugging in the Mellin space representation of the mode function, with some algebra, the term that is neither exponentially suppressed nor zero as $\tau_0 \rightarrow 0$ takes the form
\begin{align}
    &\int_{-\infty}^{\tau_0} d\tau \int_{c-i\infty}^{c+i\infty} \prod_{j=1}^4 \frac{ds_j}{2\pi i} \Gamma(s_j - \frac{\nu_j}{2}) \Gamma(s_j + \frac{\nu_j}{2}) \; C(k_1,k_2,k_3,k_4, s_1, s_2, s_3, s_4, \tau) \nonumber \\
    \times & \frac{-i}{2\pi} e^{\pi\nu}\Gamma(i \nu - n_5)\Gamma(- i \nu - n_6)\left( \frac{i\tau k_5}{2} \right)^{2n_5+2n_6} \frac{\frac{d}{2}-2(s_1+s_2-n_5) + i\nu}{\big(\frac{d}{2}-2(s_1+s_2-n_5)\big)^2 + \nu^2}\frac{(-1)^{n_5+n_6}}{n_5! n_6!} \ .
\end{align}
Expanded in large $\nu$, the first few terms of the second line look like 
\begin{equation} \label{eq:Series_1/nu_4pt_dS}
    \frac{1}{\nu^2} + \frac{-\alpha^2 - k_5^2 \tau^2}{\nu^4} + \frac{(\alpha^2 + k_5^2 \tau^2)^2 +4(1+\alpha)k_5^2 \tau^2}{\nu^6} + \cdots 
\end{equation}

On the other hand, the contact diagram calculation coming from the effective action is 
\begin{align} \label{eq:4pt_contact_dS_EFT}
    \langle \phi^4 \rangle \supset& \text{Re} \bigg( i \frac{\lambda^2}{2H^2} f_1^*(\tau_0) f_2^*(\tau_0) f_3^*(\tau_0) f_4^*(\tau_0)  \\
    &\times\int_{-\infty}^{\tau_0} \frac{d\tau}{H^4 \tau^4 } \; f_3(\tau) f_4(\tau) \frac{1}{\tau^2 \partial_\tau^2 - \tau (d-1) \partial_\tau + \nu^2 + \frac{d^2}{4} + \tau^2 k_5^2} f_1(\tau) f_2(\tau) \bigg) + ``12 \leftrightarrow 34"  \nonumber \ .
\end{align}
To compare with the previous result, notice that when acted on $f_1(\tau)f_2(\tau)$, in Mellin space, we can make the replacement 
\begin{equation}
    \left(\tau^2 \frac{d^2}{d\tau^2} - \tau (d-1) \frac{d}{d\tau} + \frac{d^2}{4}\right) (\tau^2 k_5^2)^{2n} \rightarrow \left(\frac{d}{2} - 2(s_1+s_2) +2n \right)^2 (\tau^2 k_5^2)^{2n} \ .
\end{equation}
Use this and expand in $1/\nu$ the second line of (\ref{eq:4pt_contact_dS_EFT}), we recover exactly the series (\ref{eq:Series_1/nu_4pt_dS}). See Appendix \ref{app:C} for a more detailed calculation. In fact, one can check that the first few terms match exactly between the EFT and the direct in-in calculation. We then claim that the EFT calculation captures all of the non-exponentially-suppressed terms in the direct in-in calculation.

As stated at the beginning of this section, this is not a surprise since the heavy fields redshift away with the exponential expansion, and it should decouple from the final result. Note that this does not mean we should simply use $\bar\sigma=0$. Instead, it is the homogeneous part of the heavy field that is redshifted to zero, and at the boundary we are left with the inhomogeneous solution 
\begin{equation}
    \sigma(\tau \rightarrow 0^-) = - \frac{\lambda}{\tau^2 \partial_\tau^2 - \tau(d-1)\partial_\tau + \frac{M^2}{H^2} + \tau^2 k_5^2 } \phi^2 \bigg|_{\tau \rightarrow 0^-} \ .
\end{equation}
We see no practical difference from in-in calculation using the EFT action derived from scattering amplitudes and one derived from matching the in-in correlators. 

This behavior is also consistent with our expectations for the exact RG argument in flat space. There are two complications in flat space for terms like Equation~(\ref{eq:exact_boundary_eg}),
\beq
 S_{\partial,{\rm int}} \supset \dot \phi(t_{0,+}) \phi(t_{0,-})^n \ .
\eeq
First, $\dot \phi_+$ and ${\bar \phi}$ don't commute. Second, this operator is non-local due to the time ordering. In the cosmological background, $\dot \varphi_+ \approx (k^2/a^2) \varphi \to 0$ and this operator is negligible. In addition, the commutator vanishes as $k^3/a^3$ and therefore plays no role in late time behavior. We saw in flat space that this term generates semi-local corrections, while in dS, these corrections vanish at least as fast as $k_i^2 a^{-2} \to 0$. This is different from the suppression coming from expanding in $1/M$ before taking $\varepsilon \rightarrow 0$, although such a choice is valid because we are dealing with a mass $M \gg H \sim \Lambda$. One should expect the same conclusion, but a detailed calculation is beyond the scope of this paper.

A similar procedure has been carried out in, for example, \cite{Tong:2017iat}, to obtain a quasi-single field model for inflation from a specific double field model with quadratic mixing. Related behavior has also been observed in \cite{Colas:2024ysu}, where it was shown that in de Sitter background the system remains pure up to exponentially suppressed terms when the environment traced out is heavy, corresponding to our simple example here. It would be interesting to investigate the case for a conformal environment using our formalism and compare with their plot. We leave this for a future work. 

We also emphasize that our calculation here does not take into account any exponentially suppressed terms, which is the main object studied in cosmological collider physics. It is possible that some subtleties coming from a mixed final state remain for these terms. However, cosmological observables are classical~\cite{Maldacena:2015bha,Martin:2017zxs,dePutter:2019xxv,Green:2020whw} and thus we do not expect even the exponentially suppressed terms to have the same non-commuting challenges as their flat space counterparts.

\section{Conclusions}\label{sec:conclusions}

Effective field theories have a straightforward implementation in the context of scattering experiments in flat space~\cite{Manohar:2018aog,Cohen:2019wxr}. Energy conservation creates a clean separation between the physical particles that can and cannot be produced kinematically. Moreover, the S-matrix is invariant under field redefinitions which significantly simplifies matching and classifying  EFTs~\cite{Cheung:2014dqa,Cheung:2017pzi,Cheung:2016drk}. Yet, there are many contexts where correlation functions, rather than S-matrix elements, are the observables of the theory. In these contexts, it is important to understand how and when our EFT intuitions need to be adjusted. This is particularly the case in cosmology, where EFT is an essential tool for both theory and data analysis~\cite{Cabass:2022avo} but lacks many or all of the simplifying aspects of flat space.

In this paper, we explored how to apply EFT in the context of in-in correlators. These observables can be defined in flat space, but are better known from their applications to cosmology~\cite{Weinberg:2005vy}. In the UV description, we characterize the correlations in terms of correlation functions in a pure (vacuum) state. However, after integrating out the short wavelength modes, the state of the system is no longer pure and is better described by a density matrix. This is, naively, in tension with the well-known expectation that the in-in correlators of a given theory can be determined from the S-matrix elements. Since the S-matrix does have a complete EFT description, it cannot be the case that EFT should be abandoned entirely.

The resolution that we explore in detail is that the additional contributions that appear in in-in correlators in flat space always take the form of semi-local terms, local terms, or field redefinitions in the EFT limit. These terms do not have an S-matrix interpretation as they never appear with a total energy pole. This was seen both in an explicit example of integrating out a heavy field, and non-perturbatively using exact RG. Instead, information that arises from the UV theory appears  as additional local and semi-local terms must be fixed by matching.

In the cosmological setting, in-in correlators are not expected to receive non-vanishing UV corrections beyond those associated with local field redefinitions. In single-field inflation, we can even fix the field redefinitions~\cite{Green:2020ebl} and no such corrections can arise. Like flat space, derivative corrections do appear, but time derivatives and gradients vanish due to the expansion of the universe and they do not alter the correlators (matching). Relatedly, the modes become classical and the commutators between fields and any non-local terms in the action vanish. This was seen directly in the same example of integrating out a massive field, now in de Sitter space. This is of practical importance, as EFT is used to calculate the statistical predictions for inflationary correlators in a wide range of models and it would be troubling if these predictions were incomplete.

This work is a step towards a deeper understanding of RG flow and decoupling in cosmological EFTs. The interpretation of many IR corrections in terms of anomalous dimensions~\cite{Marolf:2010zp,Hogervorst:2021uvp,DiPietro:2021sjt} points to the idea that these IR logs should be resummed using RG~\cite{Burgess:2009bs,Gorbenko:2019rza,Baumgart:2019clc,Mirbabayi:2019qtx,Green:2020txs,Cohen:2020php,Mirbabayi:2020vyt,Cohen:2021fzf}. Despite the strong evidence, the validity of this procedure non-perturbatively has not been directly proven. A sharper understanding of RG in the context of in-in correlators is a stepping stone to an all-orders understanding of the structure of EFTs and RG flows in curved and/or dynamical backgrounds.

\paragraph{Acknowledgments}
The authors thank Noah Bittermann, Tim Cohen, Thomas Colas, Kshitij Gupta, Yiwen Huang, Austin Joyce, Julio Parra-Martinez, Akhil Premkumar, and Yi Wang for helpful discussions. We are supported by the US~Department of Energy under grant~\mbox{DE-SC0009919}. 

\newpage
% -----------------------------------------------------------------------------------------------------------------------------------------
\appendix

%%%%%%%%%%%%%%%%%

\section{Well-Defined Variational Problem}\label{app:A}
In the main text we briefly mentioned that in order to have a well-defined variational problem, the Lagrangian density needs to be accompanied by some specific boundary terms. Here we will give a more detailed argument following mostly \cite{Bittermann:2022nfh, Dyer:2008hb}, and show how it works at the order $1/M^6$ in our example. 

Starting with a Lagrangian density $\L=\L(q, \dot q, \cdots, q^{(n)})$, the variational principle leads to a generalized Euler-Lagrange equation of motion 
\begin{equation}
    \frac{\partial \L }{\partial q^i} - \frac{d}{dt} \frac{\partial \L}{\partial \dot q^i} + \cdots + (-1)^n \frac{d^n}{dt^n} \frac{\partial L}{\partial (q^{(n)})^i} = 0 \ .
\end{equation}
This can be rewritten as 
\begin{equation}
    \frac{\partial^2 \L}{\partial (q^{(n)})^i \partial (q^{(n)})^j} (q^{(2n)})^j = F^i(q, \dot q, \cdots, q^{(2n-1)}) \ .
\end{equation}
In order to write this as an ODE of order $2n$, we want the Lagrangian to be non-degenerate, which in this context means 
\begin{equation}
    \frac{\partial^2 \L}{\partial (q^{(n)})^i \partial (q^{(n)})^j}  \neq 0 \ .
\end{equation}
In case the Lagrangian is not already given in this form, one can use integration by parts to massage it until it is manifestly non-degenerate, while keeping explicit the boundary terms involved. For example, suppose in a Lagrangian every single term has four or less time derivatives, then a term like $q q^{(4)}$ should be massaged into $\ddot q \ddot q$. Written in this way, a Lagrangian $\L(q, \cdots, q^{(n)})$ corresponds to an ODE of order $2n$, and we are free to impose $2n$ initial conditions. 

With this in hand, we can proceed to write the variation of the action as 
\begin{equation}
    \delta S = \int_{t_i}^{t_f} dt \; \left( \frac{\partial \L }{\partial q^i} - \frac{d}{dt} \frac{\partial \L}{\partial \dot q^i} + \cdots + (-1)^n \frac{d^n}{dt^n} \frac{\partial \L}{\partial (q^{(n)})^i} \right) \delta q + (p_1 \delta q + p_2 \delta \dot q + \cdots + p_n \delta q^{(n-1)})\bigg|_{t_i}^{t_f} \ , 
\end{equation}
where the conjugate momenta are found to be, for $m\leq n$, 
\begin{equation}
    p_m = \frac{\partial \L}{\partial q^{(m)}} - \frac{d}{dt} \left( \frac{\partial \L}{\partial q^{(m+1)}} - \cdots \frac{d}{dt}\left( \frac{\partial \L}{\partial q^{(n-1)}} - \frac{d}{dt}\frac{\partial \L}{\partial q^{(n)}} \right) \right) \ ,
\end{equation}
and the $2n$ boundary conditions can be set to $\delta q |^{t_f}_{t_i} = \delta \dot q |^{t_f}_{t_i} = \cdots = \delta q^{(n)} |^{t_f}_{t_i} = 0$. This is one choice of a well-defined variational problem, where the degrees of freedom are encoded in $q$, $\dot q$, ..., $q^{(n-1)}$. More generally, one can choose between different sets of variables as the degrees of freedom. For example, in our application we want to set $\delta \dot \phi$ free. We then replace $\delta \dot \phi$ with some other variables as the degree of freedom. In our example below we will simply guess the result. One can refer to \cite{Dyer:2008hb} for details of a more rigorous derivation. 

Let us now take a look at the $1/M^6$ term, 
\begin{equation}
    S_{EFT} \supset \int d^4x \; \lambda^2 \phi^2 \frac{\Box^2}{2M^6} \phi^2 = \int d^4x \; \lambda^2 \phi^2 \frac{\partial_t^4 - 2 \partial_t^2 \nabla^2 + \nabla^4}{2M^6} \phi^2 \ .
\end{equation}
The operator $\nabla^4$ simply translates to $k_5^4$, while the operator $\partial_t^2 \nabla^2$ can be seen as $k_5^2 \partial_t^2$, and we have already seen how to deal with $\phi^2 \partial_t^2 \phi^2$ in the main text. Their combined contribution in the in-in calculation is given by
\begin{equation}
    \lambda^2 \frac{2 k_{12}k_{34} k_5^2 + k_5^4}{2k_1 k_2 k_3 k_4 k_T M^6} \ .
\end{equation}

In the rest of this appendix we focus on the $\partial_t^4$ operator. We first write 
\begin{equation}
    \int dt \; \lambda^2 \phi^2 \frac{\partial_t^4 }{2M^6} \phi^2 = \int dt\; \frac{\lambda^2}{2M^6} (\partial_t^2\phi^2)  (\partial_t^2 \phi^2) + \frac{\lambda^2}{2M^6} \phi^2  (\partial_t^3 \phi^2) \bigg|_{t_i}^{t_f} - \frac{\lambda^2}{2M^6} (\partial_t\phi^2)  (\partial_t^2 \phi^2) \bigg|_{t_i}^{t_f}  \ .
\end{equation}
From now on we omit the $\int d^3x$ integral. Take the variation of the bulk part, we have 
\begin{align}
    &\delta\left(  \int dt \; \frac{\lambda^2}{2M^6} (\partial_t^2\phi^2)  (\partial_t^2 \phi^2) \right) \nonumber \\
    =& \int dt \; \frac{\lambda^2}{2M^6} \left(  -4\partial_t(\partial_t \phi \partial_t^2 \phi^2) + 2 \partial_t^2 (\phi \partial_t^2 \phi^2)  \right)\delta \phi \nonumber \\
    +& \frac{\lambda^2}{2M^6} \left( \big(4\partial_t^2 \phi^2 \partial_t \phi - 2 \partial_t (\phi \partial_t^2 \phi^2) \big) \delta \phi + 2 \phi \partial_t^2 \phi^2  \delta \dot \phi \right) \bigg|_{t_i}^{t_f} \ .
\end{align}
We see that, as expected, there is a term proportional to $\delta \dot \phi$ on the boundaries. However, here we would like to set $\delta \dot \phi$ free, to be consistent with the procedure when dealing with the $1/M^4$ term. Alternatively, this could be understood as setting free the initial velocity of the field $\phi$. In any case, in order to replace the $\delta \dot \phi$ term with something else, one can add to the original bulk part a boundary correction of the form $\partial_t \phi^2 \partial_t^2 \phi^2 $ to cancel out the $\delta \dot \phi$ dependence. However, it is not easy to guess the correct form of the additional boundary term from the current expression.

To get around this problem, one can regard the bulk part as a function of $\partial_t^2 {\phi^2}$. In this new variable, we would like to set $\partial_t \delta \phi^2$ free, to be consistent with the procedures in dealing with the $1/M^4$ term. That is, we would like to have a variational problem in terms of $\delta \phi^2$ and $\partial_t^2 \delta \phi^2$, instead of $\partial_t \delta \phi^2$. In this case, the boundary condition is given by $\partial_t^2 \delta \phi^2 = 0$. In terms of the original variable, $\delta \dot \phi$, this means that on the boundaries we choose $\delta \ddot \phi$ such that $2\dot \phi \delta \dot \phi + \phi \delta \ddot \phi=0$, given $\delta \phi = 0$. To do so, we need to add a boundary term to the bulk, 
\begin{align}
    &\delta\left(  \int dt \; \frac{\lambda^2}{2M^6} (\partial_t^2\phi^2)  (\partial_t^2 \phi^2) - \frac{\lambda^2}{M^6} (\partial_t\phi^2)  (\partial_t^2 \phi^2) \bigg|_{t_i}^{t_f} \right) \nonumber \\
    =& \int dt \; \frac{\lambda^2}{M^6} (\partial_t^4 \phi^2) \delta \phi^2 - \frac{\lambda^2}{M^6} \left( (\partial_t^3 \phi^2) \delta \phi^2 + (\partial_t^2 \phi^2) (\partial_t^2 \delta \phi^2) \right)\bigg|_{t_i}^{t_f} \ .
\end{align}
Now the variations are on the variables that we desire. Starting from the original EFT expression, we see the desired boundary corrections are given by 
\begin{align}
    &\int dt \; \frac{\lambda^2}{2M^6} (\partial_t^2\phi^2)  (\partial_t^2 \phi^2) - \frac{\lambda^2}{M^6} (\partial_t\phi^2)  (\partial_t^2 \phi^2) \bigg|_{t_i}^{t_f} \nonumber \\
    =& \int dt \; \lambda^2 \phi^2 \frac{\partial_t^4 }{2M^6} \phi^2 - \frac{\lambda^2}{2M^6} \phi^2  (\partial_t^3 \phi^2) \bigg|_{t_i}^{t_f} - \frac{\lambda^2}{2M^6} (\partial_t\phi^2)  (\partial_t^2 \phi^2) \bigg|_{t_i}^{t_f} \ .
\end{align}
In terms of in-in calculation, this is a contribution of the form
\begin{align}
    &\frac{4\times \lambda^2}{M^6 \times 16k_1 k_2 k_3 k_4} \left( \frac{k_{12}^4 + k_{34}^4}{k_T} -  k_{12}^3 -  k_{34}^3 -  k_{12}k_{34}^2 -  k_{12}^2 k_{34}  \right) \nonumber \\
    =&-\frac{\lambda^2}{2 k_1 k_2 k_3 k_4 k_T M^6} ( k_{12}^3 k_{34} + k_{12}^2 k_{34}^2 + k_{12}k_{34}^3 ) \ .
\end{align}
Adding the contribution from the rest of the $1/M^6$ terms, we have 
\begin{equation}
    \lambda^2 \frac{2 k_{12}k_{34} k_5^2 + k_5^4}{2k_1 k_2 k_3 k_4 k_T M^6}-\lambda^2\frac{ ( k_{12}^3 k_{34} + k_{12}^2 k_{34}^2 + k_{12}k_{34}^3 )}{2 k_1 k_2 k_3 k_4 k_T M^6}  = \lambda^2 \frac{(k_5^2 + k_{12}k_{34})^2 - k_{12}k_{34} k_T^2}{2k_1 k_2 k_3 k_4 k_T M^6} \ ,
\end{equation}
exactly what is expected from the full calculation. 

\section{Exact RG}\label{app:B}

In this appendix, we provide a more detailed derivation of the boundary terms that arise in the exact RG, following the procedure described in~\cite{Goldman:2024cvx}. We write the derivative of the partition function, defined in Equation~(\ref{eq:ZLJ}), as 
\begin{equation}
    \frac{d}{d\log \Lambda} Z[\Lambda,J]  = \int \mathcal{D}\phi \; i \frac{d}{d\log \Lambda} T \left( S_0 + S_{I}  \right) e^{iS} \ ,
\end{equation}
where $S_{I} = S_{\partial} + S_{\rm int}$ includes both a bulk interaction term $S_{\rm int}$ and a boundary term $S_\partial$ that is localized at $t=t_0$. Schematically, the idea is to write
\bea
\frac{d}{d\log \Lambda} S_0 &=&\frac{d}{d\log \Lambda} K^{-1} \left( \frac{1}{2}|\varphi^2|(t_0)+\frac{1}{2} \int_\gamma dt   \partial_\mu \varphi  \partial^\mu \varphi \right) \\
&=& \frac{1}{2} \frac{d}{d\log \Lambda} K^{-1}  \int_\gamma dt_1\int_\gamma dt_2 \varphi(t_1)\Box G(t_1,t_2) \Box \varphi(t_2) + {\rm B.T} \  \\
&=& - \frac{1}{2} \frac{d}{d\log \Lambda} K \frac{\delta}{\delta \varphi} S_0 \star G \star \frac{\delta}{\delta \varphi} S_0 \ ,
\eea
where ${\rm B.T}$ is a term at $t_0$ we will derive later and $\star$ indicates that we are integrating over time. Here $G(t,t')$ is the path ordered propagator, which satisfies 
\begin{equation}
    (\Box - m^2)_x G(x,y) = (\Box - m^2)_y G(x,y)= \delta^{(4)}(x-y) \ .
\end{equation}
Using the Schwinger-Dyson equations, we can exchange $S_0$ for $S_I$ such that
\begin{equation}
   \mathcal{O} \frac{\delta S_0}{\delta \phi}  =  \frac{\delta \mathcal{O}}{\delta \phi} - \mathcal{O} \frac{\delta S_{I}}{\delta \phi}   \ .
\end{equation}
This equation should be understood to hold inside the path integral as a result of integration by parts. We can apply the Schwinger-Dyson equations twice to get
\begin{align}
-\frac{1}{2} \frac{\delta S_0}{\delta \varphi} \star G \star \frac{\delta S_0}{\delta \varphi} & =-\frac{1}{2} \int_\gamma dt \left(\frac{\delta^2 S_0}{\delta \varphi \delta \varphi} G(t,t)\right)+\frac{1}{2} \frac{\delta S_0}{\delta \varphi} \star G \star \frac{\delta S_I}{\delta \varphi} \\
& =-\frac{1}{2} \int_\gamma dt\left(\frac{\delta^2 S_0}{\delta \varphi \delta \varphi} G(t,t)\right)+\frac{1}{2} \int_\gamma dt \left(\frac{\delta^2 S_I}{\delta \varphi \delta \varphi}  G(t,t)\right)-\frac{1}{2} \frac{\delta S_I}{\delta \varphi} \star G \star \frac{\delta S_I}{\delta \varphi} \ . \nonumber
\end{align}
Therefore, if we define
\beq
\frac{d}{d\log \Lambda} S_I = \frac{d}{d\log \Lambda} K \left(\frac{1}{2} \frac{\delta S_I}{\delta \varphi} \star G \star \frac{\delta S_I}{\delta \varphi} - \frac{1}{2} \int_\gamma dt \left(\frac{\delta^2 S_I}{\delta \varphi \delta \varphi}  G(t,t)\right) \right) + {\rm B.T.} \ ,
\eeq
the additional constant term can be absorbed into the overall normalization of the partition function,
\beq
\delta \log Z \supset -\frac{1}{2} \int_\gamma dt\left(\frac{\delta^2 S_0}{\delta \varphi \delta \varphi} G(t,t)\right) \ .
\eeq
This could also be absorbed into the action as constant, but from either perspective it has no impact on the dynamics of $\varphi$. With these changes, the partition function is indeed independent of $\Lambda$,
\beq
\frac{d}{d\log \Lambda}  Z[J,\Lambda] = 0 \ .
\eeq
This argument is essentially Polchinski's original derivation. However, we have yet to to determine (a) which terms appear in $S_\partial$ and $S_{\rm int}$, and (b) if the action remains local after integrating out.

In order to understand the challenge, we should note that the time-ordering along the contour has greatly simplified our derivation, at the cost of obscuring the subtleties. To illustrate, we start with the term 
\beq
\frac{\delta S_I}{\delta \varphi} \star G \star \frac{\delta S_I}{\delta \varphi} =\int_\gamma dt_1  \int_\gamma dt_2 \frac{\delta S_I}{\delta \varphi(t_1)} G(t_1,t_2) \frac{\delta S_I}{\delta \varphi(t_2)} \ .
\eeq
Notice that the integrals both run over the entire contour. In contrast, the term that we would derive if we simply substituted the in-out expression for the effective action would be
\beq
S_{\rm EFT} =\int^{t_0}_{-\infty(1+ i\epsilon)} dt {\cal L}_{\rm EFT}  \supset \int^{t_0}_{-\infty(1+ i\epsilon)} dt_1  \int^{t_0}_{-\infty(1+ i\epsilon)} dt_2 \frac{\delta S_I}{\delta \varphi(t_1)} G(t_1,t_2) \frac{\delta S_I}{\delta \varphi(t_2)}  \ .
\eeq
The conventional terms therefore do not capture the operators
\begin{align}
\frac{\delta S_I}{\delta \varphi} \star G \star \frac{\delta S_I}{\delta \varphi} \supset& \int^{t_0}_{-\infty(1+ i\epsilon)} dt_1  \int^{t_0}_{-\infty(1- i\epsilon)}dt_2 \frac{\delta S_I}{\delta \varphi(t_1)} G(t_1,t_2) \frac{\delta S_I}{\delta \varphi(t_2)} 
\\
&+ \int^{t_0}_{-\infty(1- i\epsilon)} dt_1  \int^{t_0}_{-\infty(1+ i\epsilon)}dt_2 \frac{\delta S_I}{\delta \varphi(t_1)} G(t_1,t_2) \frac{\delta S_I}{\delta \varphi(t_2)} + {\rm B.T.} \ \label{eq:Exact_step1}
\end{align}
These terms are particularly significant, as we have a source at $t_0$, $J(\x) \varphi(\x,t_0)$. The operators above are non-local with respect to the time ordering, and therefore we cannot simply interpret them as terms in a local action.

In order to understand the additional terms that are not captured by $S_{\rm EFT}$, we will have to be more careful to split the contours into regions defined by $t_\pm - t_0 = t (1\pm i \epsilon)$ and $\varphi_\pm = \varphi(t_\pm)$. We also recall that the path ordering is such that $t_-$ is time-ordered and $t_+$ is anti-time ordered. We will additionally define $\varphi(t_0) = {\bar \varphi}$ to distinguish terms that are purely local at $t_0$ and those that arise from the time evolution. The time integrals along the contour then become
$$
\int_{-\infty_-}^{0} dt + \int_{0}^{-\infty_+} dt =  \int_{-\infty_-}^{0} dt - \int_{-\infty_+}^{0} dt = \int_\gamma dt \ , 
$$
In terms of these variables, the partition function of the theory becomes
\beq
    Z = \int \mathcal{D} \varphi \; e^{iS[\phi]} =\int \D {\bar \varphi} \int^{{\bar \varphi}} \D \varphi_+  \int^{{\bar \varphi}} \D \varphi_-  \; e^{iS[\varphi_+; \varphi_-] + I[{\bar \varphi}, J]} \ ,
\eeq
where $I[{\bar \varphi},J]$ is a purely local function of the boundary field, $\bar\varphi(\x)$, and its source, $J(\x)$. As a result, the action along the contour (ignoring the local terms in ${\bar \varphi}$) becomes
\bea
    S_0[\varphi_\pm] &=& \int d^4x \left( -\frac{1}{2} \partial_\mu \varphi \partial^\mu \varphi - \frac{1}{2} m^2 \varphi^2 \right) K^{-1} \nonumber \\
    &=& \int d^4x \left( \frac{1}{2} \varphi (\Box - m^2) \varphi - \frac{1}{2} \partial_\mu (\varphi \partial^\mu \varphi)  \right)K^{-1}\\
     &=& \int d^4x \left( \frac{1}{2} \varphi (\Box - m^2) \varphi - \frac{1}{2} \partial_\mu (\varphi \partial^\mu \varphi)  \right)K^{-1} + \int d^3 x  \frac{1}{2}\left( {\bar \varphi} \dot \varphi_- - {\bar \varphi} \dot \varphi_+ \right) K^{-1}  \ . \nonumber
\eea
Here we are using $K^{-1}$ abstractly, as we defined it in Fourier space. Note that the second term, which also appears in the on-shell wavefunction of the free theory, is not purely local on the boundary because $\dot \varphi_\pm$ is discontinuous at $t_\pm =t_0$.

In order to identify additional boundary terms that may arise from the derivation of the exact RG, we can write the action 
\beq
\bar S_0 \equiv S_0 -  \int d^3 x  \frac{1}{2}\left( {\bar \varphi} \dot \varphi_- - {\bar \varphi} \dot \varphi_+ \right) K^{-1}  = \int d^4x  \frac{1}{2} \varphi (\Box - m^2) \varphi K^{-1} \ ,
\eeq
which can be written in terms of the propagator as 
\begin{equation}
    \bar S_0 = \int d^4x d^4y \; \frac{1}{2} \varphi(x) G^{-1}(x,y)\varphi(y) K^{-1} \ . 
\end{equation}
From here, we can repeat the argument above, keeping track of the use of integration by parts, with the goal of writing
\begin{equation}
    \frac{d}{d\log \Lambda} \bar S_0 = - \int d^4x d^4y \; \frac{1}{2} \frac{\delta \bar S_0}{\delta \varphi(x)} G(x,y) \frac{\delta \bar S_0}{\delta \varphi(y)} \frac{d}{d\log \Lambda} K + {\rm B.T.} 
\end{equation}
Again, we note that in this process the expression above still has mixed $t_+$ and $t_-$ terms from the integration over the full contours. For the moment, our goal is to identify the precise form of the boundary terms that appear at this stage. In momentum space, the propagator $G(\k,t,t')$ is local in $\k$, and therefore one can focus only on the time coordinate for a single $\k$ value, with the understanding that all formulas should include a $\int d^3k / (2\pi)^3$. By construction, we know that 
\beq
\int_\gamma dt dt' \varphi(t) (\Box-m^2)_t G(t,t') (\Box-m^2)_t'\varphi(t') = \int_\gamma dt \varphi(t) (\Box-m^2) \varphi(t) \ . \label{eq:bt1}
\eeq
In addition, we know that 
\beq
\frac{\delta \bar S_0}{\delta \varphi(t) } = (\Box -m^2) \varphi(t) \ .
\eeq
Integrating Equation~(\ref{eq:bt1}) by parts twice yields
\begin{align}
-\frac{1}{2} \int_\gamma dt \varphi(t) (\Box-m^2) \varphi(t) = &- \int d^4x d^4y \; \frac{1}{2} \frac{ \delta \bar S_0}{\delta \varphi(x)} G(x,y) \frac{\delta \bar S_0}{\delta \varphi(y)} \frac{d}{d\log \Lambda} K \\
-\frac{1}{2} \int_\gamma dt'  \frac{d}{d\log \Lambda} K^{-1} \Bigg({\bar \varphi} \partial_t G(t,t')|^{t=t_{0,+}}_{t=t_{0,-}}&-(\dot \varphi_+ - \dot \varphi_- )|_{t=t_0} G(t_0,t')\Bigg)(\Box-m^2)_{t'}\varphi(t') \ .
\end{align}
At this stage, the result is not a boundary term as we still have a $t'$ integral. However, we can now integrate by parts with respect to $t'$ to find that the second term is
\begin{align}
&-\frac{1}{2} \int_\gamma dt'  \frac{d}{d\log \Lambda} K^{-1} \Bigg({\bar \varphi} \partial_t G(t,t')|^{t=t_{0,+}}_{t=t_{0,-}}-(\dot \varphi_+ - \dot \varphi_- )|_{t=t_0} G(t_0,t')\Bigg)(\Box-m^2)_{t'}\varphi(t')  \\ 
=&-\frac{1}{2}  \frac{d}{d\log \Lambda} K^{-1}  \Bigg({\bar \varphi} \partial_t G(t,t_0)|^{t=t_{0,+}}_{t=t_{0,-}}(\dot \varphi_+ - \dot \varphi_- )-(\dot \varphi_+ - \dot \varphi_- ) G(t_0,t_0)(\dot \varphi_+ - \dot \varphi_- ) \\
& -{\bar \varphi}^2 \partial_t \partial_{t'} G(t,t')|^{t=t_{0,+}}_{t=t_{0,-}}|^{t'=t_{0,+}}_{t'=t_{0,-}}  +(\dot \varphi_+ - \dot \varphi_- )|_{t=t_0} \partial_{t'} G(t_0,t')|^{t'=t_{0,+}}_{t'=t_{0,-}}{\bar \varphi} \Bigg) \ .
\end{align}
The two additional terms that arise from equations acting on $G(t,t')$ cancel. One can then check that 
\beq
\partial_t \partial_{t'} G(t,t')|^{t=t_{0,+}}_{t=t_{0,-}}|^{t'=t_{0,+}}_{t'=t_{0,-}}= 0 \ ,
\eeq
and therefore our final expression for the boundary term is
\bea
{\rm B.T.} &=& -\frac{1}{2}  \frac{d}{d\log \Lambda} K^{-1}  \Bigg( -G(t_0,t_0) (\dot \varphi_+ - \dot \varphi_- )^2 \\
&&+{\bar \varphi} \partial_t G(t,t_0)|^{t=t_{0,+}}_{t=t_{0,-}}(\dot \varphi_+ - \dot \varphi_- )+ (\dot \varphi_+ - \dot \varphi_- )|_{t=t_0} \partial_{t'} G(t_0,t')|^{t'=t_{0,+}}_{t'=t_{0,-}}{\bar \varphi}  \Bigg)  \ .
\eea
One can check that this result is the same as in~\cite{Goldman:2024cvx} restricted to $\varphi_+(t_0) =\varphi_-(t_0)={\bar \varphi}$, the diagonal elements of the density matrix.

We see that this term is local on the boundary. However, due to the time derivatives, the ordering of the operators at $t_0$ still matters since $[\dot \varphi_\pm, J\varphi(t_0)] \neq 0$ for a generic $J(\k)$. In spite of this, because the boundary term here is purely quadratic, if we only consider $J(\k)$ with vanishing overlap with $\partial_{\log \Lambda} K^{-1}$, the operator ordering is irrelevant. This is the case relevant to Section~\ref{sec:ERG} and therefore we can ignore the quadratic boundary terms in the main text.

%%%%%%%%%%%%%%%%%%%%%%%%%%%%%%%%%%%%%%%%%%%%%%%%%%%
\section{Mellin Space Trispectrum from in-in and EFT}\label{app:C}
%%%%%%%%%%%%%%%%%%%%%%%%%%%%%%%%%%%%%%%%%%%%%%%%%%%
In this appendix, we display a detailed calculation of the trispectrum in de Sitter spacetime using Mellin representation, and compare between the exchange diagram calculated using the full theory and the EFT. 

In terms of the mode function, our goal is to calculate 
\begin{align}
    & 2\text{Re} \left( -\lambda^2 \int_{\infty}^{\tau_0} \frac{d\tau_2}{H^4\tau_2^4} \int_{\infty}^{\tau_2} \frac{d\tau_1}{H^4\tau_1^4} \langle \phi^2 \sigma (\tau_1) \phi^2 \sigma(\tau_2) \phi^4(\tau_0) - \phi^2 \sigma (\tau_1) \phi^4(\tau_0) \phi^2 \sigma(\tau_2) \rangle \right)  \\
    =& 2\text{Re} \bigg( -\lambda^2 f_1^*(\tau_0)f_2^*(\tau_0)f_3^*(\tau_0)f_4^*(\tau_0) \int_{\infty}^{\tau_0} \frac{d\tau_2}{H^4\tau_2^4} \int_{\infty}^{\tau_2} \frac{d\tau_1}{H^4\tau_1^4} f_1(\tau_1)f_2(\tau_1)f_3(\tau_2)f_4(\tau_2)g_5(\tau_1)g_5^*(\tau_2) \nonumber \\
    &+\lambda^2 f_1^*(\tau_0)f_2^*(\tau_0)f_3(\tau_0)f_4(\tau_0) \int_{\infty}^{\tau_0} \frac{d\tau_2}{H^4\tau_2^4} \int_{\infty}^{\tau_2} \frac{d\tau_1}{H^4\tau_1^4} f_1(\tau_1)f_2(\tau_1)f_3^*(\tau_2)f_4^*(\tau_2)g_5(\tau_1)g_5^*(\tau_2)\bigg) \ .\nonumber
\end{align}
The second term factorizes into two parts that look similar to the three point function. We will discuss it at the end of this appendix. The first term, however, does not factorize. The mode functions, $f_k$ and $g_k$, are determined in terms of the Mellin representation of the Hankel functions
\begin{align}\label{eq:mode_function_dS}
    & H\sqrt{\frac{\pi}{4}}e^{-\frac{i\pi}{4}(1+2\nu)} (-\tau)^{d/2} H_{\nu}^{(2)}(-k\tau)\nonumber \\
    =& \frac{i}{\pi} H \sqrt{\frac{\pi}{4}} e^{-\frac{i\pi}{4}} (-\tau)^{d/2} \int_{c-i\infty}^{c+i\infty} \frac{ds}{2\pi i} \Gamma(s - \frac{\nu}{2}) \Gamma(s + \frac{\nu}{2}) (-\frac{ik\tau}{2})^{-2s} \ .
\end{align}
We can also write the mode function in the limit $k\tau_0 \rightarrow 0$ using the following relation,
\begin{equation}\label{eq:H_nu^1_zero_asymp}
    H_{\nu_j}^{(1)} (-k_j \tau_0) \xrightarrow{k\tau_0 \rightarrow 0} -\frac{i}{\pi} \left( \Gamma(\nu_j)\left( - \frac{k_j \tau_0}{2} \right)^{-\nu_j} + e^{-i\pi\nu_i}\Gamma(-\nu_j)\left( - \frac{k_j \tau_0}{2} \right)^{\nu_j} \right) \ .
\end{equation}
We see that we need to evaluate the real part of
\begin{align}
    -\sum_{{\color{blue} \bar\nu_i} = \pm \nu_i}& \lambda^2 H^2 \frac{2}{(4\pi)^5} \Gamma({\color{blue} \bar\nu_1}) \Gamma({\color{blue} \bar\nu_2}) \Gamma({\color{blue} \bar\nu_3}) \Gamma({\color{blue} \bar\nu_4}) e^{\frac{i}{2}\pi ({\color{blue} \bar\nu_1 + \bar\nu_2 + \bar\nu_3 + \bar\nu_4})}(-\tau_0)^{6-({\color{blue} \bar\nu_1 + \bar\nu_2 + \bar\nu_3 + \bar\nu_4})} k_1^{{\color{blue} - \bar\nu_1}} k_2^{{\color{blue} - \bar\nu_2}} k_3^{{\color{blue} - \bar\nu_3}} k_4^{{\color{blue} - \bar\nu_4}} \nonumber \\
    \times & 2^{\color{blue} \bar\nu_1 + \bar\nu_2 + \bar\nu_3 + \bar\nu_4} \int_{-\infty}^{\tau_0} d \tau_2 \int_{-\infty}^{\tau_2} d \tau_1 \int_{c-i\infty}^{c+i\infty} \prod_{j=1}^4 \frac{ds_j}{2\pi i} \Gamma(s_j - \frac{\nu_j}{2}) \Gamma(s_j + \frac{\nu_j}{2}) \; i^{-2\sum_{j=1}^6 s_j} 2^{2\sum_{j=1}^6 s_j} \nonumber \\
    \times & k_1^{-2s_1} k_2^{-2s_2} k_3^{-2s_3} k_4^{-2s_4} k_5^{-2s_5 -2s_6} \Gamma(s_5 - \frac{\nu_5}{2}) \Gamma(s_5 + \frac{\nu_5}{2}) \Gamma(s_6 - \frac{\nu_5^*}{2}) \Gamma(s_6 + \frac{\nu_5^*}{2}) e^{2s_6 i\pi} \nonumber \\
    \times & (-\tau_1)^{\frac{d}{2} -1 - 2(s_1 + s_2 + s_5)} (-\tau_2)^{\frac{d}{2} -1 - 2(s_3 + s_4 + s_6)} \ .
\end{align}
Here we use $\color{blue}\bar\nu_i$ and $\nu_i$ to represent different kinds of $\nu$'s. The $\color{blue} \bar\nu_i$'s come from the mode functions with $k\tau_0 \rightarrow 0$, and each one can take either plus or minus sign, corresponding to the two terms in (\ref{eq:H_nu^1_zero_asymp}). At the end we sum over the different choices of the $\color{blue}\bar\nu_i$'s. To make our life easier, let us write $\nu_5 = i \nu$, where $\nu$ is a real number. This could be done because $\nu_5$ is purely imaginary for very heavy $M$. We could continue to carry out the time integral and sum up all the relevant poles. However, as our goal is to compare with the EFT calculation, we would take a different path, and make the comparison rather than the explicit result manifest. Let us start with this question: since now integrals over $s_5$ and $s_6$ pick up poles that are complex ($s_{5/6} \rightarrow  \pm i\nu - n_{5/6}$, where $n_{5/6}$ are non negative integers), what happens if their imaginary parts do not cancel? In addition to oscillating terms like $(-k_5 \tau_0)^{\pm 2 i \nu}$, there will be factors that look like 
\begin{equation}
    \Gamma(\pm i\nu -n_5)\Gamma(\pm i\nu -n_6) e^{i\pi( n_5 - n_6)} \ .
\end{equation}
If we take the limit $\nu \rightarrow \infty$, this leads to an exponential suppression $e^{-\pi \nu}$, multiplied by some polynomial factors made up of $n_5$, $n_6$ and $1/\nu$. Therefore, such terms correspond to the exponentially suppressed (in $\nu$, or $M$) parts of the final result, which cannot be captured by the EFT prescription. On the other hand, if the imaginary parts cancel, the same factor above becomes 
\begin{equation}
    \Gamma(\pm i\nu -n_5)\Gamma(\mp i\nu -n_6) e^{\pm\pi\nu + i\pi( n_5 - n_6)} \ .
\end{equation}
We see that now we have an extra factor of $e^{\pi\nu}$ to treat the exponential suppression. Therefore, we will focus on the poles where the imaginary part of $s_5$ and $s_6$ cancel. 

With this in mind, we update the formula by taking $s_5 = \pm i\nu/2 - n_5$ and $s_6 = \mp i\nu/2 - n_6$, 
\begin{align}\label{eq:mellin_4pt_exchange}
    -\sum_{{\color{blue} \bar\nu_i} = \pm \nu_i}& \lambda^2 H^2 \frac{2}{(4\pi)^5} \Gamma({\color{blue} \bar\nu_1}) \Gamma({\color{blue} \bar\nu_2}) \Gamma({\color{blue} \bar\nu_3}) \Gamma({\color{blue} \bar\nu_4}) e^{\frac{i}{2}\pi ({\color{blue} \bar\nu_1 + \bar\nu_2 + \bar\nu_3 + \bar\nu_4})}(-\tau_0)^{6-({\color{blue} \bar\nu_1 + \bar\nu_2 + \bar\nu_3 + \bar\nu_4})} k_1^{{\color{blue} - \bar\nu_1}} k_2^{{\color{blue} - \bar\nu_2}} k_3^{{\color{blue} - \bar\nu_3}} k_4^{{\color{blue} - \bar\nu_4}} \nonumber \\
    \times & 2^{\color{blue} \bar\nu_1 + \bar\nu_2 + \bar\nu_3 + \bar\nu_4} \sum_{n_5,n_6}^{\infty} \int_{-\infty}^{\tau_0} d \tau_2  \int_{c-i\infty}^{c+i\infty} \prod_{j=1}^4 \frac{ds_j}{2\pi i} \Gamma(s_j - \frac{\nu_j}{2}) \Gamma(s_j + \frac{\nu_j}{2}) \; i^{-2\sum_{j=1}^4 s_j + 2 n_5 - 2 n_6 \mp 2i\nu} \nonumber \\
    \times & \Gamma(\pm i\nu - n_5) \Gamma(\mp i\nu - n_6) k_1^{-2s_1} k_2^{-2s_2} k_3^{-2s_3} k_4^{-2s_4} k_5^{2n_5 + 2n_6}   \frac{-(-\tau_2)^{d -1 - 2(s_1 + s_2 + s_3 + s_4 - n_5 - n_6) }}{\frac{d}{2}-2(s_1 + s_2 \pm \frac{i\nu}{2} - n_5)} \nonumber \\
    \times &2^{2\sum_{j=1}^4 s_j - 2n_5 - 2n_6}\frac{(-1)^{n_5+n_6}}{n_5! n_6!} \ ,
\end{align}
where we have carried out the $\tau_1$ integral. We can further rewrite the denominator as 
\begin{equation}
    \frac{1}{\frac{d}{2}-2(s_1+s_2-n_5) \mp i\nu} = \frac{\frac{d}{2}-2(s_1+s_2-n_5) \pm i\nu}{\big(\frac{d}{2}-2(s_1+s_2-n_5)\big)^2 + \nu^2} \ ,
\end{equation}
and collect all the terms involving $\nu$, $n_5$ and $n_6$. Choosing the term in which we have a factor of $e^{\pi\nu}$ to treat the exponential suppression, this gives us 
\begin{align}
    \sum_{{\color{blue} \bar\nu_i} = \pm \nu_i}& i\lambda^2 H^2 \frac{1}{(4\pi)^4} \Gamma({\color{blue} \bar\nu_1}) \Gamma({\color{blue} \bar\nu_2}) \Gamma({\color{blue} \bar\nu_3}) \Gamma({\color{blue} \bar\nu_4}) (-\tau_0)^{6-({\color{blue} \bar\nu_1 + \bar\nu_2 + \bar\nu_3 + \bar\nu_4})} k_1^{{\color{blue} - \bar\nu_1}} k_2^{{\color{blue} - \bar\nu_2}} k_3^{{\color{blue} - \bar\nu_3}} k_4^{{\color{blue} - \bar\nu_4}} \nonumber \\
    \times &  \int_{-\infty}^{\tau_0} d \tau_2  \int_{c-i\infty}^{c+i\infty} \prod_{j=1}^4 \frac{ds_j}{2\pi i} \Gamma(s_j - \frac{\nu_j}{2}) \Gamma(s_j + \frac{\nu_j}{2}) \; i^{-2\sum_{j=1}^4 s_j +(\color{blue} \bar\nu_1 + \bar\nu_2 + \bar\nu_3 + \bar\nu_4) } \nonumber \\
    \times &  k_1^{-2s_1} k_2^{-2s_2} k_3^{-2s_3} k_4^{-2s_4} (-\tau_2)^{d -1 - 2(s_1 + s_2 + s_3 + s_4) } 2^{2\sum_{j=1}^4 s_j - (\color{blue} \bar\nu_1 + \bar\nu_2 + \bar\nu_3 + \bar\nu_4) }\\
    \times& \sum_{n_5,n_6}^{\infty} \frac{-i}{2\pi} e^{\pi\nu}\Gamma(i \nu - n_5)\Gamma(- i \nu - n_6)\left( \frac{i\tau_2 k_5}{2} \right)^{2n_5+2n_6} \frac{\frac{d}{2}-2(s_1+s_2-n_5) + i\nu}{\big(\frac{d}{2}-2(s_1+s_2-n_5)\big)^2 + \nu^2}\frac{(-1)^{n_5+n_6}}{n_5! n_6!}  .
\end{align}
We then single out the last line which includes all the $n_5$, $n_6$ and $\nu$ dependencies, 
\begin{equation}
    \sum_{n_5,n_6}^{\infty} \frac{-i}{2\pi} e^{\pi\nu}\Gamma(i \nu - n_5)\Gamma(- i \nu - n_6)\left( \frac{i\tau k_5}{2} \right)^{2n_5+2n_6} \frac{\frac{d}{2}-2(s_1+s_2-n_5) + i\nu}{\big(\frac{d}{2}-2(s_1+s_2-n_5)\big)^2 + \nu^2}\frac{(-1)^{n_5+n_6}}{n_5! n_6!} \ .
\end{equation}
The factor of $1/2\pi$ is included for later convenience. We also included a phase factor $-i$. This comes from the fact that in order for the final result to be independent of $\tau_0$ (so that it does not redshift away at $\tau_0 \rightarrow 0$), we have $9 - 2(s_1+s_2+s_3+s_4-n_5-n_6)-({\color{blue} \bar\nu_1 + \bar\nu_2 + \bar\nu_3 + \bar\nu_4}) = 0$, which implies that the total power of $i$ should be given by $i^{-9-2i\nu - 4n_6 + 2({\color{blue} \bar\nu_1 + \bar\nu_2 + \bar\nu_3 + \bar\nu_4})}$. With $n_6$ being some integer, this is simply $-ie^{\pi\nu}i^{2({\color{blue} \bar\nu_1 + \bar\nu_2 + \bar\nu_3 + \bar\nu_4})}$, hence the factor of $-i$. In other words, the phase for the first three lines is simply $i^{2({\color{blue} \bar\nu_1 + \bar\nu_2 + \bar\nu_3 + \bar\nu_4})}$.

In our case, $\nu_1 = \nu_2 = \nu_3 = \nu_4 = 3/2$. As a result, if all the $\color{blue}\bar\nu_i$'s are positive, we get a factor of $i^{12} = 1$. If we flip a sign of one of the $\color{blue}\bar\nu_i$'s, we will pick up an extra factor of $i^{-6} = -1$, which does not swap the role of the real and imaginary parts. At the same time, other factors change with the different choice of the $\color{blue}\bar\nu_i$, so that the different terms do not cancel. As a result, as far as taking the real part is concerned, the terms in the full expression can only differ by an extra minus sign multiplied by some real factors, depending on the choices of the signs of the $\color{blue}\bar\nu_i$'s.

Noticing that $e^{\pi\nu}\Gamma(i \nu - n_5)\Gamma(- i \nu - n_6)$ in general is of order $1/\nu^{n_5+n_6+1}$ or higher, we can write out the lowest few orders of the expansion in $1/\nu$ after summing over the first few $n_5$ and $n_6$. The first few (real) terms are
\begin{equation}{\label{eq:gamma_n5n6_expansion}}
    \frac{1}{\nu^2} + \frac{-\alpha^2 - k_5^2 \tau^2}{\nu^4} + \frac{(\alpha^2 + k_5^2 \tau^2)^2 +4(1+\alpha)k_5^2 \tau^2}{\nu^6} + \cdots 
\end{equation}
where $\alpha = \frac{d}{2} - 2(s_1 + s_2)$. 
How does this compare with the EFT calculation? First, in this example, the EFT interaction term is given by 
\begin{equation}\label{eq:L_int_dS_EFT}
    \L_I = \frac{\lambda^2}{2 H^2}\phi^2 \frac{1}{\tau^2 \partial_\tau^2 - \tau (d-1) \partial_\tau + \frac{M^2}{H^2} + \tau^2 k_5^2} \phi^2 \ .
\end{equation}
As a result, the four point contact term, in Mellin space, takes the form 
\begin{align}
    \sum_{{\color{blue} \bar\nu_i} = \pm \nu_i}& i\lambda^2 H^2 \left( \frac{1}{4\pi} \right)^4 \Gamma({\color{blue}\bar\nu_1}) \Gamma({\color{blue}\bar\nu_2}) \Gamma({\color{blue}\bar\nu_3}) \Gamma({\color{blue}\bar\nu_4}) (-\tau_0)^{6} k_1^{{\color{blue} - \bar\nu_1}} k_2^{{\color{blue} - \bar\nu_2}} k_3^{{\color{blue} - \bar\nu_3}} k_4^{{\color{blue} - \bar\nu_4}} \left( - \frac{i}{2 \tau_0} \right)^{{\color{blue} \bar\nu_1 + \bar\nu_2 + \bar\nu_3 + \bar\nu_4}} \nonumber \\
    \times& \int_{c-i\infty}^{c+i\infty} \prod_{j=1}^4 \frac{ds_j}{2\pi i} \Gamma(s_j - \frac{\nu_j}{2}) \Gamma(s_j + \frac{\nu_j}{2})  i^{-2(s_1 + s_2 + s_3 + s_4) } 2^{2(s_1 + s_2 + s_3 + s_3) } k_1^{-2s_1} k_2^{-2s_2} k_3^{-2s_3} k_4^{-2s_4}   \nonumber \\
    \times&\int_{-\infty}^{\tau_0} d\tau \; (-\tau)^{ - 1 - 2( s_3 + s_4)  }\frac{1}{\tau^2 \partial_\tau^2 - \tau (d-1) \partial_\tau + \frac{M^2}{H^2} + \tau^2 k_5^2} (-\tau)^{d-2(s_1+s_2)} \ . %%%%%%%%%%%%%  
    \label{eq:mellin_eft_trispec}
\end{align}
To make further progress, notice that time derivatives can be exchanged with factors of $s_i$'s. 
For example, looking at the form of the mode function (\ref{eq:mode_function_dS}), we see that when acting on the fields in Mellin space, we can replace $\tau\partial_\tau \rightarrow \frac{d}{2} - 2 s$. When acting on $\phi^2$, we have
\begin{equation}
    \left[ \tau^2 \partial^2_\tau - (d-1)\tau \partial_\tau \right] (k\tau)^{2n}  \rightarrow \left[\left( \frac{d}{2} - 2 (s_1+s_2-n) \right)^2 - \left( \frac{d}{2} \right)^2 \right] (k\tau)^{2n} \ .
\end{equation}
This implies 
\begin{equation}
    \tau^2 \partial_\tau^2 - \tau (d-1) \partial_\tau + \frac{M^2}{H^2} \rightarrow \left(\frac{d}{2}-2(s_1+s_2)\right)^2 + \nu^2 = \alpha^2 + \nu^2 \ .
\end{equation}
One might be tempted to simply carry out this replacement in (\ref{eq:L_int_dS_EFT}). However, expanded in terms of large $M$, (\ref{eq:L_int_dS_EFT}) will involve terms like
\begin{equation}
    \left(\tau^2 \frac{d^2}{d\tau^2} - \tau (d-1) \frac{d}{d\tau} + \frac{d^2}{4}\right) (\tau^2 k_5^2)^{2n} \rightarrow \left(\frac{d}{2} - 2(s_1+s_2) +2n \right)^2 (\tau^2 k_5^2)^{2n} \ .
\end{equation}
Carefully treating the extra $\tau k_5$ factors, up to order $1/\nu^6$, we have 
\begin{align}
    &\frac{1}{\tau^2 \partial_\tau^2 - \tau (d-1) \partial_\tau + \frac{M^2}{H^2} + \tau^2 k_5^2}\nonumber \\
    =&\frac{1}{\nu^2}\left[1 - \frac{\tau^2 \partial_\tau^2 - \tau (d-1) \partial_\tau + \frac{d^2}{4} + \tau^2 k_5^2}{\nu^2} + \frac{\left(\tau^2 \partial_\tau^2 - \tau (d-1) \partial_\tau + \frac{d^2}{4} + \tau^2 k_5^2\right)^2}{\nu^4} - \cdots \right]\nonumber \\
    \rightarrow & \frac{1}{\nu^2}\left[ 1 - \frac{\alpha^2 + k_5^2\tau^2}{\nu^2} + \frac{\alpha^4 + k_5^4\tau^4 + k_5^2\tau^2 \alpha^2 + (\alpha + 2)^2 k_5^2\tau^2 }{\nu^6} - \cdots \right] \nonumber \\
    =&\frac{1}{\nu^2} - \frac{\alpha^2 + k_5^2 \tau^2}{\nu^4} + \frac{(\alpha^2 + k_5^2 \tau^2)^2 +4(1+\alpha)k_5^2 \tau^2}{\nu^6} - \cdots 
\end{align}
We see that it takes the same form as (\ref{eq:gamma_n5n6_expansion}). Plugging this back into (\ref{eq:mellin_eft_trispec}), we recover exactly (\ref{eq:mellin_4pt_exchange}). One can check that the next few orders in $1/\nu$ from the two expansions also match. We then claim that indeed the EFT calculation matches the full result in this case. 

Lastly, we check what happens to the factorized term. It turns out, that part is completely exponentially suppressed. To see this, notice that the factorized integrals are 
\begin{equation}
    f_1^*(\tau_0)f_2^*(\tau_0)\int_{-\infty}^{\tau_0} \frac{d\tau_1}{H^4\tau_1^4} f_1(\tau_1)f_2(\tau_1) g_5(\tau_1),\;\; f_3(\tau_0)f_4(\tau_0) \int_{-\infty}^{\tau_0} \frac{d\tau_2}{H^4\tau_2^4} f_3^*(\tau_2)f_4^*(\tau_2)g_5^*(\tau_2) \ .
\end{equation}
These are almost the same as the three point contact integrals, but without a $g_5(\tau_0)$ (or its complex conjugate). This makes it 
\begin{align}
    \sum_{{\color{blue} \bar\nu_i} = \pm\nu_i}&\left( \frac{1}{4\pi} \right)^3 (-\tau_0)^{6/2 {\color{blue} - \bar\nu_1 - \bar\nu_2}} \Gamma({\color{blue} \bar\nu_1}) \Gamma({\color{blue} \bar\nu_2}) k_1^{{\color{blue} -\bar\nu_1}} k_2^{{\color{blue}-\bar\nu_2}} 2^{\color{blue}\bar\nu_1 + \bar\nu_2 } e^{\frac{i}{2}\pi ({\color{blue} \bar\nu_1 + \bar\nu_2 })} \nonumber \\
    \times& \int_{c-i\infty}^{c+i\infty} \frac{ds_1}{2\pi i} \frac{ds_2}{2\pi i} \frac{ds_5}{2\pi i}\; \frac{-(-\tau_0)^{\frac{d}{2} - 2(s_1 + s_2 + s_5) }}{\frac{d}{2} - 2(s_1 + s_2 + s_5)} e^{-i\pi(s_1 + s_2 + s_5) } 2^{2(s_1 + s_2 + s_5)} k_1^{-2s_1} k_2^{-2s_2} k_3^{-2s_5} \nonumber \\
    \times & \Gamma(s_1 - \frac{\nu_1}{2}) \Gamma(s_1 + \frac{\nu_1}{2}) \Gamma(s_2 - \frac{\nu_2}{2}) \Gamma(s_2 + \frac{\nu_2}{2}) \Gamma(s_5 - \frac{i\nu}{2}) \Gamma(s_5 + \frac{i\nu}{2}) \ .
\end{align}
Now, if we take the pole $\frac{d}{2} - 2(s_1 + s_2 + s_5)$, we will have exponential suppression coming from $\Gamma(\frac{d}{4}-s_1-s_2 - \frac{i\nu}{2}) \Gamma(\frac{d}{4}-s_1-s_2 + \frac{i\nu}{2})$, but there is no additional exponential factor to treat it. As a result, this term is exponentially suppressed, and we do not expect it to be captured by the EFT. 

\clearpage
\phantomsection
\addcontentsline{toc}{section}{References}
\small
\bibliographystyle{utphys}
\bibliography{Refs}

\end{document}